\documentclass[superscriptaddress,amsmath,amssymb, aps, prx,longbibliography,twocolumn, footinbib]{revtex4-2}

\usepackage{upgreek}
\usepackage{color}
\usepackage{graphicx}
\usepackage{nccmath}
\usepackage{bm}
\usepackage{hyperref}
\hypersetup{colorlinks,breaklinks,
            urlcolor=[rgb]{0,0,0.36},
            linkcolor=[rgb]{0,0,0.36},
            citecolor=[rgb]{0,0,0.36}}
\usepackage[mathlines]{lineno}
\usepackage{dcolumn}
\usepackage{mathtools}  
\usepackage{siunitx}    % SI unit abbreviations
\usepackage{dsfont}
\usepackage{extarrows}
\usepackage{epsfig,tabularx,multirow,booktabs}
\usepackage{braket}
\usepackage{chemformula}

\usepackage{mystyle}
\usepackage[utf8]{inputenc}

\newcommand{\Harvard}{Department of Physics, Harvard University, Cambridge, MA 02138, USA.}

\newcommand{\WM}{Department of Physics, University of Wisconsin-Madison, Madison, Wisconsin 53706, USA}

\newcommand{\ETH}{Institute for Theoretical Physics, ETH Zurich, 8093 Zurich, Switzerland.}

\newcommand{\Geneva}{Department of Quantum Matter Physics, University of Geneva,
24 quai Ernest-Ansermet, 1211 Geneva, Switzerland.}

\begin{document}

\title{Local noise spectroscopy of  Wigner crystals in two-dimensional materials}

\author{Pavel~E.~Dolgirev}
\thanks{P.E.D. and I.E. contributed equally to this work.}
\affiliation{\Harvard}
\author{Ilya~Esterlis}
\thanks{P.E.D. and I.E. contributed equally to this work.}
\affiliation{\WM}
\affiliation{\Harvard}
\author{Alexander~A.~Zibrov}
\affiliation{\Harvard}
\author{Mikhail~D.~Lukin}
\affiliation{\Harvard}
\author{Thierry~Giamarchi}
\affiliation{\Geneva}
\author{Eugene~Demler}%
\affiliation{\ETH}

\date{\today}

\begin{abstract}
We propose to use local electromagnetic noise spectroscopy as a 
versatile and noninvasive tool to study Wigner crystal phases of strongly-interacting two-dimensional electronic systems. 
In-plane imaging of the local noise is predicted to enable single-site resolution of the electron crystal when the sample-probe distance is less than the inter-electron separation. At larger sample-probe distances, noise spectroscopy encodes information about the low-energy Wigner crystal phonons, including the dispersion of the transverse shear mode, the pinning resonance due to disorder, and optical modes emerging, for instance, in bilayer crystals.
We  discuss the potential utility of local noise probes in analyzing the rich set of phenomena expected to occur in the vicinity of the melting transition.
\end{abstract}

\maketitle

Wigner crystal (WC) phases of the electron gas have been a subject of active research since their initial conception by Wigner many years ago~\cite{wigner_crystal}. Recently, a new generation of experiments providing compelling evidence of WC phases across a number of two-dimensional electron gas (2DEG) systems \cite{Smolenski_2021, Zhou_2021, Hossain_2020, Hossain_2021, Falson_2022} have reinvigorated interest in the field for a number of reasons: i) The experiments are carried out at low temperatures in the degenerate regime $T \ll E_F$ ($E_F$ is the Fermi energy) and at zero perpendicular magnetic field, ii) observation of
 unexpected and potentially exotic magnetism in the vicinity of the WC melting transition~\cite{Hossain_2020, Hossain_2021, Falson_2022}, iii) in the case of WCs in transition metal dichalcogenide (TMD) systems, optical spectroscopy enabled direct measurement of the WC ordering wave vector~\cite{Smolenski_2021}, and iv) TMD bilayer WCs appear stable up to anomalously high electron densities and temperatures~\cite{Zhou_2021}.
 
Despite both novel and improved experimental capabilities for clarifying the onset of crystallization, there remain few probes for characterizing salient properties of the WC phase~\footnote{A notable new probe is the STM imaging technique developed in \cite{Li:2021}.}. These include 
the nano- and meso-scale structure of the electron crystal, as well as properties of the low-energy WC phonons~\cite{kukushkin1994evidence}. 
The necessity of experimental proposals is especially pressing in the TMD systems, for which many conventional measurements, such as transport, are not possible due to notorious challenges associated with large contact resistances \cite{cui_2015, cui_2017}. 

In the present paper we propose local electromagnetic noise spectroscopy as a probe of WC states, and consider the conditions under which such measurements are within current experimental reach. We demonstrate that, owing to the large emergent length scale associated with the WC lattice constant, magnetic noise spectroscopy can be used to both map local charge properties at the WC lattice scale, as well as long-wavelength properties of the WC phonons. We demonstrate that magnetic noise sensing is especially well-suited to probe a defining feature of the WC solid -- the transverse shear mode. Other resonances unique to the crystal phase can also be observed, such as pinning of the WC by disorder and optical modes in more complex crystals, such as bilayer WCs. Our proposal is in part inspired by developments in the field of ``qubit" sensors, in which quantum impurities of various sorts are used to probe local electromagnetic fields and their associated fluctuations. Notable probes include nitrogen-vacancy (NV) and silicon-vacancy (SiV) centers in diamond~\cite{hong2013nanoscale,grinolds2013nanoscale,rondin2014magnetometry,shields2015efficient,kolkowitz2015probing,dovzhenko2018magnetostatic,casola2018probing,hsieh2019imaging,andersen2019electron,thiel2019probing,rustagi2020sensing,chatterjee2021semiconductor,zhou2021magnon,zhang2021ac,wang2022noninvasive,wang2023visualization}, hBN defects~\cite{gottscholl2021room,gottscholl2021spin,castelletto2021color,huang2022wide,vaidya2023quantum,healey2023quantum}, and SNOM detectors~\cite{bonnell2012imaging,jiang2016generalized}, which can sense magnetic and/or electric fields~\cite{dolde2014nanoscale,myers2017double,wolfowicz2018electrometry,yang2020vector,bian2021nanoscale,qiu2022nanoscale}.

\begin{figure*}[t!]
    \centering
    \includegraphics[width=\textwidth]{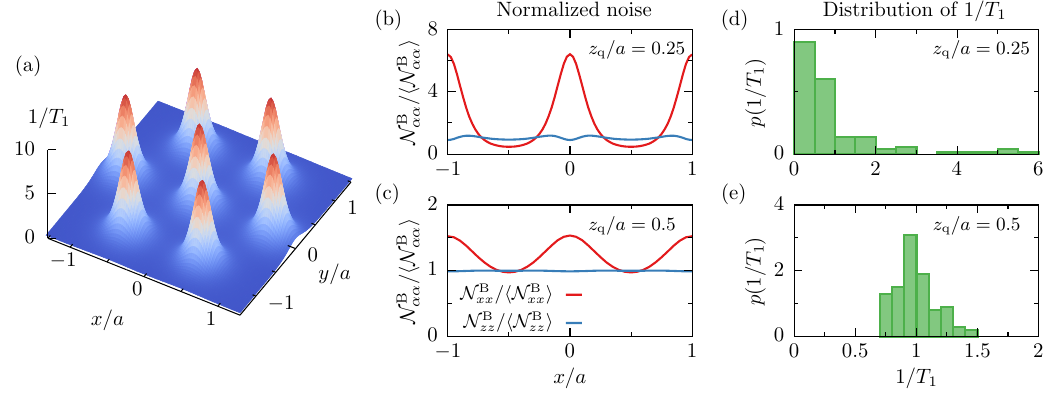}
    \caption{Single-site resolution (SSR) of the WC with local magnetic noise spectroscopy. (a) Spatial dependence of the $1/T_1$ relaxation rate of the qubit probe showing that the magnetic noise is strongly enhanced when the qubit is placed on top of an electron site. Here we fixed $z_{\rm q} = 0.25 a$, the qubit quantization axis is aligned with $\hat{z}$, and we used the phonon Green's function of the clean WC -- see Appendix~\ref{appendix:SER}. Panels (b) and (c) represent cuts of ${\cal N}_{xx}^{\rm B}$ and ${\cal N}_{zz}^{\rm B}$ along one of the edges in the triangular WC, showing that i) the local magnetic noise is strongly anisotropic and ii) when the probe is further away from the sample, the noise it senses appears more homogeneous. This is further illustrated in panels (d) and (e), where the broad distribution of $1/T_1$ at $z_{\rm q} = 0.25 a$ (d) becomes notably narrower at $z_{\rm q} = 0.5 a$ (e).     Various quantities are normalized by their spatial averages, $\langle\ldots\rangle$, to highlight the magnitude of spatial fluctuations in the SSR regime.  
    }
    \label{fig:SSR}
\end{figure*}

We consider an atomic scale qubit probe that is brought near a 2D sample of interest and is used to sense local magnetic fields. Fluctuating currents in the sample generate stray magnetic fields, which then affect the relaxation properties of the qubit~\cite{Langsjoen:2012}. These relaxation properties are directly related to the magnetic noise tensor:
    \begin{align}
        {\cal N}^{\rm B}_{\alpha\beta}(\bm r_{\rm q}, \omega) & =\frac{1}{2} \int \dd t ~ e^{i\omega t}\langle  \{B_\alpha(\bm r_{\rm q}, t), B_\beta(\bm r_{\rm q},0)\}\rangle_T, \label{eq:NB}
    \end{align}
where $\{.,.\}$ is the anticommutator, $\langle \ldots \rangle_T$ denotes the thermal expectation value, $\bm r_{\rm q} = (\bm r,z_{\rm q})$ is the position of the noise probe, and Greek indices $\alpha$, $\beta$, etc., denote Cartesian components.  The Biot-Savart law further relates the magnetic field to currents via $B_\alpha(\bm q,z) = (2\pi / qc) e^{-q|z|}\epsilon_{\alpha\beta\gamma}(iq_\beta - q \delta_{\beta z})j_\gamma(\bm q) \equiv K_{\alpha\gamma}(\bm q, z)j_\gamma(\bm q)$, with $j_\alpha(\bm q)$ being the 2D current density in the sample. Here $\bm q$ is the in-plane wave vector, $c$ is the speed of light, and $\epsilon_{\alpha\beta\gamma}$ is the Levi-Civita tensor \footnote{As a concrete example, in the case of an NV center, the dynamic magnetic noise would, depending on the direction of magnetic field, cause transitions between the states with $m_s=0$ and $m_s=\pm 1$, or lead to fluctuations of energy differences between states with different $m_s$ \cite{casola2018probing}. In Appendix~\ref{appendix:Electric noise}, we also analyze the electrical noise that can cause transitions between $m_s=-1$ and $m_s=+1$, resulting in energy difference fluctuations between the states with $m_s=0$ and $|m_s|=1$ and, thus, affecting $1/T_2$-like measurements~\cite{myers2017double}.}. Experimentally, one usually accesses ${\cal N}^{\rm B}_{\alpha\beta}$ via $1/T_1$ relaxometry and/or $1/T_2$ spin-echo-like measurements, which are related to the noise via Fermi's golden rule. The dependence of the magnetic noise tensor on various physical parameters -- such as temperature, frequency, qubit's position, electron density, etc. -- allows one to characterize intrinsic correlations of the 2D material~\cite{Agarwal2017Magnetic,casola2018probing,rodriguez2018probing,flebus2018quantum,chatterjee2019diagnosing,dolgirev2022characterizing,chatterjee2022single,machado2022quantum}.

The key idea of our work is based on the observation that the WC lattice constant, which is tuned by varying the electron density, can be made much larger than the underlying microscopic lattice scale of the 2D material. 
For instance, in TMDs the WC lattice constant varies in the range $a\simeq 10$-$30\,$nm~\cite{Zhou_2021, Smolenski_2021}. This opens the possibility that a qubit probe can be brought closer to the 2D sample than the inter-electron distance, allowing for spatial resolution of magnetic noise produced by individual electrons in the WC. As will be elaborated upon below, magnetic noise in a WC is sourced primarily by charges oscillating about their equilibrium lattice sites, that is, by local phonon fluctuations. In the regime $z_{\rm q} \lesssim a$, the noise sensor effectively probes the local phonon density of states $g(\omega)$, with the noise directly on top of an electron being approximately given by (c.f. Eq.~\eqref{eq:NBr})
    \be
    \mathcal N^{\rm B}(\omega) \sim \frac{T n e^2}{c^2 m} g(\omega)\Big(\sum_\bfG e^{-z_{\rm q} G}\Big)^2,
    \ee
where $\bfG$ are reciprocal lattice vectors of the WC and $m$ is the effective electron mass.
In TMD systems specifically, the noise will receive an enhancement owing to the relatively large melting temperatures (on the order of tens of K) and the relatively high electron densities ($n\sim 10^{11}$-$10^{12} \,\text{cm}^{-2}$). The magnitude of the density of states contribution depends on the ratio of the probe frequency to the plasma frequency, $\omega_p^2 = 2\pi n e^2/m a$, which is the characteristic phonon frequency scale in a WC. 
For WCs realized in TMD systems, $\omega_p$ is on the order of a few THz. Typical resonant frequencies of atom-like solid-state defects, however, are in the  GHz range, so we expect $\omega \ll \omega_p$. The density of states then comes primarily from low-frequency transverse phonons, $g(\omega) \approx \omega / 2\pi v_s^2$, where $v_s$ is the sound speed. In this regime, we estimate the noise will be within the sensitivity of current sensors (roughly $\mathcal N^{\rm B} \gtrsim 1\, \text{pT}^2 \times \text{Hz}^{-1}$) for a sample-probe distance on the order of a few nm. For probes with operating frequencies in the THz, such as tin-vacancy sensors~\cite{guo2023microwave} and SNOM detectors~\cite{cocker2021nanoscale}, the density of states contribution would be larger and the sample-probe distance could be tens of nm. Additionally, nonlinear optics methods may be used to push the qubit operating regime to higher frequency~\cite{ludovic2018nitrogen,wang2022sensing}. Our estimates indicate that while single-site resolution (SSR) of the WC using noise spectroscopy is challenging, it  nevertheless, can be within experimental reach. In what follows, we develop the general microscopic theory for electromagnetic noise from a WC.  In addition to the SSR regime, we will also show that the noise allows one to study long-wavelength WC phonons when $z_{\rm q}\gtrsim a$.

Deep in the WC phase, current fluctuations are generated by the time-varying polarization from fluctuating WC phonons, which are the primary 
low-energy degrees of freedom in the system. 
The phonon spectrum is described by the elastic potential energy $U_{\rm el} = \frac 12 \sum_{\bm q} \sum_{\lambda}  m \omega^2_\lambda(\bm q) |u_\lambda (\bm q)|^2.$ Here $\lambda$ is the phonon mode index, which includes the transverse (shear) and longitudinal (compression) modes, as well as optical modes in crystals with more than one electron per unit cell (such as in the bilayer WC); $u_\lambda(\bm q)$ are the associated phonon displacements with in-plane wave vector $\bm q$; and the mode frequencies are $\omega_\lambda(\bm q)$. These mode frequencies take into account both the Coulomb forces between electrons, as well as the effects of weak disorder -- a point we will elaborate upon below. The response properties of the WC are encoded in the phonon Green's function:
    \begin{align}
    D_{\alpha\beta}(\bm q, \omega) &= -i \int_0^\infty \dd t ~ e^{i\omega t}\langle [u_\alpha(\bm q, t), u_\beta(-\bm q,0)]\rangle_T.
    \end{align} 
In terms of more familiar quantities, the phonon Green's function can be directly related to the non-local optical conductivity of the WC~\cite{brem2022terahertz} -- see Appendix~\ref{appendix:sigma+D}. 

The fluctuation-dissipation theorem, together with the Bio-Savart law, relates the magnetic noise tensor to the current-current response function of the electron system. The latter can be expressed using the phonon Green’s function to yield:
\begin{widetext}
\begin{align}
{\cal N}^{\rm B}_{\alpha\beta} & (\bm r, \omega) \approx  2T n e^2\omega\, \text{Im}\Big[ \sum_{\bfG_1,\bfG_2} e^{i\bm r\cdot(\bfG_1 - \bfG_2)} \int_{\rm 1BZ}\frac{\dd^2\bm q}{(2\pi)^2} ~ {\cal K}_{\alpha\gamma}(\bm q + \bfG_1, z_{\rm q}){\cal K}_{\beta\delta}(-\bm q - \bfG_2, z_{\rm q}) D_{\gamma\delta}(\bm q,\omega)
\Big].\label{eq:NBr}
\end{align}
\end{widetext}
Here the $\bm q$-integration is over the first Brillouin (1BZ) of the WC lattice, $e$ is the electron charge, and we have assumed $T\gg \hbar \omega$. The exponential $z$-dependence of the Biot-Savart kernel $\mathcal K_{\alpha\beta}$ implies that the qubit probe effectively averages over a spatial region with a size determined by $z_{\rm q}$ 
and, thus, one should distinguish two regimes: $z_{\rm q} \lesssim a$ and $z_{\rm q} \gtrsim a$.

In the limit $z_{\rm q} \lesssim a$ (SSR regime), the contribution of non-zero $\bfG$'s, which determine the intra-unit cell structure, is important. In this case, the largest contribution to the noise comes from positions $\bm r$ near the WC lattice sites, where the oscillating phase factors in Eq.~\eqref{eq:NBr} go to one. In Fig.~\ref{fig:SSR} we demonstrate the spatial dependence of the qubit $1/T_1$ (which is simply related to particular components of the magnetic noise tensor -- see Appendix~\ref{appendix:SER}). Figure \ref{fig:SSR} illustrates the important characteristics of the noise in the SSR regime: i) strong spatial inhomogeneity, ii) anisotropy of the noise tensor, and iii) broad distribution of qubit relaxation rates from different points in the plane. These characteristic features become weaker upon increasing the qubit-sample distance $z_{\rm q}$, as seen in Fig.~\ref{fig:SSR}c,e. The anisotropy of the noise near an electron site can be understood from the fact that an oscillating dipole in the plane would emit primarily in the direction perpendicular to the plane. 

In the opposite limit $z_{\rm q} \gtrsim a$, the noise \eqref{eq:NBr} is well approximated by keeping only the $\bfG = 0$ terms and simplifies to
\begin{equation}
    {\mathcal N}^{\rm B}_{zz}(\omega)  \approx \frac{\pi T ne^2 \omega}{c^2z_{\rm q}^2} \int_0^\infty \dd x \,  x \, e^{-x} \, \mathrm{Im}\Big[D_{\rm T}\Big(\frac{x}{2z_{\rm q}},\omega\Big)\Big].
   \label{eq:NB2}
\end{equation}
Notably, in this limit the noise becomes independent of the in-plane position $\bm r$. In obtaining this result we have utilized that for $qa \ll 1$ and $(\omega z_{\rm q}/c)^2\ll 1$, one can approximate ${\cal N}^{\rm B}_{\alpha\beta}(\omega) \approx \text{diag}({\cal N}^{\rm B}_{zz}(\omega)/2, {\cal N}^{\rm B}_{zz}(\omega)/2 ,{\cal N}^{\rm B}_{zz}(\omega) )$, that is, there is only one independent component of the noise tensor (see Appendix~\ref{appendix:Magnetic noise}).
We also used the fact that, at long-wavelengths $qa\ll 1$, the phonon Green's function may be decomposed into transverse (T) and longitudinal (L) parts~\cite{bonsall_elastic_wigner}. We observe that ${\mathcal N}^{\rm B}_{zz}(\omega)$ is determined by the transverse phonon Green's function, $D_{\rm T}$, which encodes 
the transverse sound mode in the system. The existence of this mode captures the hallmark feature of the crystal phase -- its rigidity to shear. In the limit of a clean WC, the dispersion of the transverse mode is $\omega_{\rm T}(\bm q) \approx v_s q$ for $qa \ll 1$.

\begin{figure}
    \centering
    \includegraphics[width=\columnwidth]{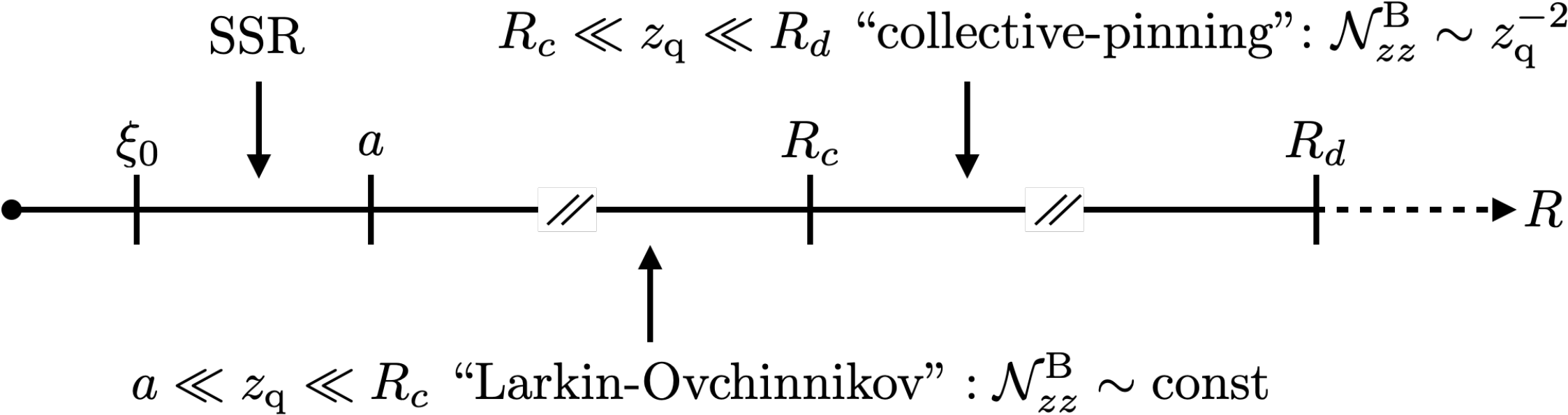}
    \caption{Hierarchy of the relevant length scales for a weakly disordered WC, as described in the main text. Beyond the length scale $R_d \gg R_c$, topological defects become important.  }
    \label{fig:length_scales}
\end{figure}

Any realistic 2DEG system is affected by inhomogeneities of the sample. Here we have implicitly assumed such disorder effects are not strong enough to completely destroy the local crystalline order of the WC. (In this regard, the SSR method would allow one to image the distorted lattice giving access to the average displacements due to disorder and long-range crystalline correlations.) While leaving the crystal intact, weak disorder nevertheless has important effects on the phonon spectrum at larger length scales, relevant when $z_{\rm q} \gg a$. The most significant effect is the ``pinning" of the crystal, which opens a (pseudo) gap in the phonon spectrum: $\omega_\lambda(q) \to \omega_0$ as $q \to 0$~\footnote{To be more precise, weak disorder gives rise to a distribution of gap frequencies, which manifests as an imaginary (frequency-dependent) correction to the $\omega_0$-pole in the phonon propagator -- see Appendix~\ref{appendix:GVM}. }.
This has important physical consequences: the pinned WC is an insulator (in the absence of pinning the WC could conduct by sliding) and there emerges a finite frequency ``pinning" resonance, $\omega_{\rm pin}$, in the absorption spectrum~\cite{fukuyama_pinning,chitra_wigner_hall,fogler_pinning_wigner, chitra_wigner_long, chitra_wigner_zerob}. In the absence of an applied magnetic field,  $\omega_{\rm pin} = \omega_0$, whereas in a large out-of-plane magnetic field, $\omega_{\rm pin} = \omega_0^2/\omega_c$, where $\omega_c = eB/mc$ is the cyclotron frequency~\footnote{We note that, to experimentally access the SSR regime, the suppression of low-energy density of states $g(\omega)$ due to development of the pinning (pseudo) gap will likely further experimentally constrain $\omega \gtrsim \omega_0$.}.

The pinning frequency defines an important characteristic length scale according to $\omega_0 \sim v_s/R_c$, where $R_c$ is known as the ``Larkin length" ~\cite{Larkin_model_1970-SovPhysJETP31_784,larkin_ovchinnikov_pinning}. The Larkin length, assumed to satisfy $R_c \gg a$, is the length scale at which electrons ``feel" the stochastic aspects of the disorder potential and metastability can manifest. Specifically, it is the length scale at which relative phonon displacements become of the order of a relevant micropscopic length, $\xi_0$, which may correspond to the width of the electronic wave function localized to the WC lattice sites, the correlation length of the disorder potential, or the magnetic length in cases with a large perpendicular magnetic field. The various important length scales are summarized in Fig.~\ref{fig:length_scales}.

Additionally, weak disorder also gives rise to broadening of the otherwise long-lived phonon modes. The foregoing discussion motivates the following simple parameterization of the phonon Green's function in the regime $qa \ll 1$:
	\be
	D_{\rm T}(q,\omega) = - \frac 1m\frac{1}{\omega^2 + 2 i \gamma \omega -  (v_s^2 q^2  + \omega_0^2)} 
	\label{eq:DT},
	\ee
where $\gamma \sim \omega_0$ is the damping rate. 
Here we have assumed that the length scales being probed are sufficiently long that the relevant observables are self-averaging, and translation invariance is effectively restored, implying in-plane momentum $\bm q$ is a good quantum number~\footnote{Strictly speaking, this analysis should be valid for $z_{\rm q}\gtrsim R_c$. For a more quantitative understanding of the regime $a\lesssim z_{\rm q} \lesssim R_c$, a different treatment of disorder effects might be required. In this regime, the probe is close enough to the sample so that there is no self-averaging yet, while, at the same time, the local properties are important. One implication could be is that in this regime, one should use the form as in Eq.~\eqref{eq:DT}, except with much smaller linewidth $\gamma$, which is expected to make the magnetic noise larger. We leave careful analysis of this regime to future work.}. 
This phenomenological form of the Green's function is in agreement with the results of more detailed calculations -- see, for instance, Appendix~\ref{appendix:GVM}, where we treat the disorder using the replica trick and Gaussian variational method (GVM) developed in Refs.~\cite{Giamarchi:1995, chitra_wigner_long, chitra_wigner_zerob, giamarchi_1998_Young321, giamarchi_quantum_pinning}.

\begin{figure}
    \centering
    \includegraphics[width=\columnwidth]{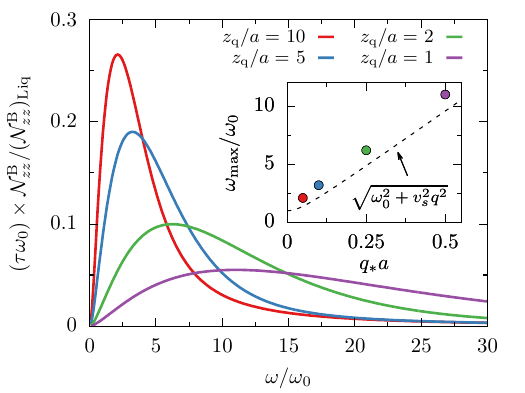}
    \caption{ Collective behavior in the WC.
    Magnetic noise as a function of frequency $\omega$ for various probe heights $z_{\rm q}$ shows an enhancement upon crossing the transverse phonon frequency $\omega = \omega_{\rm T}(q_*)$ at $q_* = 1/2z_{\rm q}$. Inset: tracking the maxima of ${\cal N}^{\rm B}_{zz}(\omega)$ for various $z_{\rm q}$ enables reconstructing the dispersion curve $\omega_{\rm T}( q)$.%, as further discussed in the main text.
    }
    \label{fig:freq_scaling}
\end{figure}

Utilizing the form of the Green's function \eqref{eq:DT}, Eq.~\eqref{eq:NB2} for the noise becomes
\be
\begin{aligned}
{\cal N}^{\rm B}_{zz}(\omega) &= \left[{\cal N}^{\rm B}_{zz}\right]_\mathrm{Liq} \frac{1}{\tau\omega_0} \int_0^\infty \dd x ~ x \, e^{-x} \\
&\qquad \times\frac{2\hat\gamma \hat\omega^2}{\{\hat\omega^2 - [\hat\omega_{\rm T}^2(x/2z_{\rm q}) + 1]\}^2+4\hat\gamma^2\hat\omega^2},
\end{aligned}
\label{eq:NB_WC}
\ee 
where $\hat\omega = \omega/\omega_0$ and $\hat\gamma = \gamma/\omega_0$. For the reference noise,
we used the Johnson–Nyquist noise in the metallic phase
$[{\cal N}^{\rm B}_{zz}]_\mathrm{Liq} = \pi T\sigma_0/(c^2 z_{\rm q}^2)$, where $\sigma_0 = ne^2\tau/m$ is the
Drude conductivity and $\tau$ is the scattering time. While the liquid state noise $[{\cal N}^{\rm B}_{zz}]_\mathrm{Liq}$ is essentially featureless as a function of $\omega$ and $z_{\rm q}$, the noise in the WC phase exhibits a much richer structure. An immediate conclusion from Eq.~\eqref{eq:NB_WC} is that the low-frequency magnetic noise in the WC phase is significantly suppressed relative to that of the liquid:  ${\cal N}^{\rm B}_{zz} \sim [{\cal N}^{\rm B}_{zz}]_\text{Liq} \times (\omega/\omega_0)^2$ as  $\omega \to 0$,  yielding a crude signature of the transition from the metallic to insulating phase \footnote{We remark  the precise power law of the noise at low frequencies is still an open question requiring a more sophisticated treatment. Despite this theoretical uncertainty, one might nevertheless attempt to do generalized echo-like measurements, in particular, analogs of the CPMG pulse sequence~\cite{machado2022quantum}, as they can give direct experimental access to the scaling of the noise with frequency.}. 
For an estimate of the liquid-state noise near the WC transition, we consider a TMD system at $T \sim 10 ~ \text K$, $z_{\rm q} \sim 10 ~ \text{nm}$, and $n \sim 10^{11} ~ \text{cm}^{-2}$, and use the mobilities reported in Ref.~\cite{Larentis:2018}. This yields $[{\cal N}^{\rm B}_{zz}]_\mathrm{Liq}\sim 5$ pT$^2 \times$Hz$^{-1}$, which is within the sensitivity of current qubit sensors~\cite{andersen2019electron,shields2015efficient}. For bilayer WCs, observed to be stable up to significantly higher densities and temperatures~\cite{Zhou_2021}, the noise will be further enhanced  (see Appendix~\ref{appendix:SER}). We thus expect detection of the WC transition with noise sensing, via both $1/T_1$- and $1/T_2$-measurements, is within experimental reach.

\begin{figure}
    \centering
    \includegraphics[width=1\columnwidth]{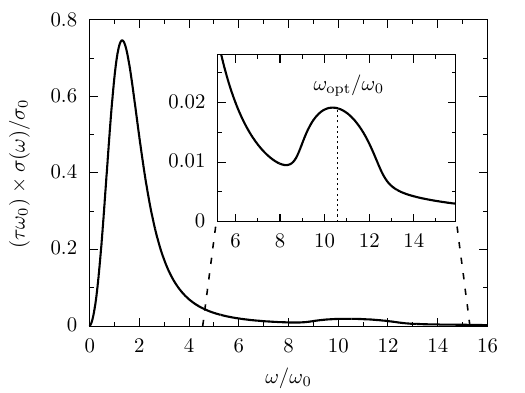}
    \caption{Conductivity in the bilayer WC exhibits an additional peak at the optical phonon frequency, which couples to the noise sensor due to the presence of a weak disorder.}
    \label{fig:sig_bilayer}
\end{figure}

More refined information may be extracted by considering how the noise varies with the probe frequency $\omega$ and height $z_{\rm q}$. As shown in  Fig.~\ref{fig:freq_scaling}, there is a resonant enhancement of the magnetic noise when $\omega \approx \sqrt{v_s^2 q_*^2 + \omega_0^2}$, where the wave-vector is determined by the probe height, $q_* = 1/2z_{\rm q} $. The peak position moves closer to the phonon dispersion as the phonons become sharper, i.e., as the disorder effects become weaker. Thus, for a sufficiently clean WC corresponding to $\omega_0 \ll \omega_p$, 
mapping of the magnetic noise in the $(z_{\rm q}, \omega)$-space allows for direct extraction of the transverse phonon dispersion curve. Importantly, even if stronger disorder precludes a straightforward mapping of the dispersion curve, at large enough probe heights (small enough wave-vectors) the noise still exhibits a resonant enhancement at $\omega_0$. Similarly, we anticipate that other $q=0$ resonances unique to WC phases can be studied with the noise measurements. One notable example is the optical phonon present in bilayer WCs, corresponding to out-of-phase charge oscillations between the layers. Without disorder, the qubit probe separated farther than the interlayer spacing simply averages over the layers and is insensitive to the optical mode. With disorder, however, differences in pinning between the layers will couple the optical mode into the layer-averaged response, as we illustrate in Fig.~\ref{fig:sig_bilayer} (see also Appendix~\ref{appendix:GVM}). We note that other interesting optical modes have also been recently predicted for WCs in multi-valley 2DEGs~\cite{calvera_2022}.

Experimental feasibility of mapping the phonon spectrum requires an estimate of the pinning frequency $\omega_0$. This frequency is determined by the disorder of the sample, making a direct evaluation from microscopic considerations challenging. However, if we assume the disorder effects are relatively weak, as  evidenced by the appreciable WC correlation length inferred from experiments~\cite{Smolenski_2021}, then it is reasonable to assume $\omega_0 \lesssim \omega_p$. To be within the operating regime of NV or SiV centers, one requires $\omega_0 \lesssim 50 \, \text{GHz}$; see also Fig.~\ref{fig:freq_scaling}. This operating regime can potentially be extended to higher frequencies via nonlinear frequency mixing methods~\cite{ludovic2018nitrogen,wang2022sensing}.
Finite temperature effects should also help push $\omega_0$ into the experimentally accessible range, as phonon frequencies are generally expected to soften upon approaching the thermal melting transition. Fabrication of cleaner TMD samples will also both increase the overall noise and decrease $\omega_0$. Application of a large perpendicular magnetic field will parameterically shift the pinning resonance to a lower frequency as well as make it narrower~\cite{chitra_wigner_hall,chitra_wigner_long, fogler_pinning_wigner}. All of these  should increase the feasibility of our proposal. We analyse the noise for the case of systems without time-reversal symmetry, such as those in a magnetic field, in Appendices~\ref{appendix:Magnetic noise} and~\ref{appendix:Electric noise}.

To summarize, the measurements we propose could be used to characterize properties of the WC at both short distances and long distances and low energies. In the short distance SSR regime, in addition to the direct imaging of the WC lattice one could potentially extract 
a number of important properties relating to the nano- and meso-scale properties of the system. This would be particularly useful to study the physics near quantum melting of the WC, where there have been proposals of intermediate phases involving meso-scale inhomogeneity and other forms of symmetry breaking such as nematicity \cite{Spivak:2004, Jamei:2005}. In the regime of a far-separated sensor, monitoring the evolution of the phonon spectrum upon increasing electron density would shed light on quantum effects in the WC. Some interesting questions in this regard include the extent to which magnetic tendencies of the WC are encoded in the elastic coefficients of the crystal and the role of phonon softening for melting. Reasonable estimates suggest local noise spectroscopy of the sort we propose is feasible for characterizing properties of the recently discovered WCs in TMD systems. 
Although we haven't explored it in detail here, the spin properties of the WC, which are expected to be particularly rich near melting~\cite{Chakravarty:1999, Kim:20222}, can also be probed via magnetic noise sensing. Beyond WCs, we also expect the techniques described here to be useful in studying moiré systems, which have a similarly large emergent length scale associated with the moiré unit cell.

\section*{ACKNOWLEDGEMENTS}
The authors would like to thank E.~Kaxiras, A.~Barr, A.~Imamoglu, P.~Volkov, C.~Kuhlenkamp, R.~Xue, J.~Curtis, Z.~Sun, M.~Fogler, V.~Falko, R. Citro, A. Yacoby, and K.~Agarwal for fruitful discussions. The work of P.E.D. was sponsored by the Army Research Office and was accomplished under Grant Number W911NF-21-1-0184. I.E. acknowledges support from the AFOSR Grant No. FA9550-21-1-0216 and the University of Wisconsin--Madison. E.D. acknowledges support from the SNSF project 200021\_212899. T.G. acknowledges support from the SNSF project 200020\_188687. M.D.L. and A.A.Z. were supported by  the NSF and CUA.

\bibliography{wc}

\clearpage
\newpage

\onecolumngrid
\appendix

\section{Relating the long-wavelength optical conductivity and the phonon Green's function}
\label{appendix:sigma+D}

In this Appendix, we derive the relationship between the non-local long-wavelength optical conductivity and the phonon Green's function. We do this in the general case of a lattice with a basis relevant for bilayer WC systems, as well in the absence of time-reversal symmetry so that the Hall conductivity $\sigma_{\rm H}$ can be nonzero. Our discussion follows closely that given in Ref.~\cite{bonsall_elastic_wigner} for the dielectric susceptibility of a WC.

Because the WC is an insulator, it is convenient to work with the polarization density (as opposed to the current density). Phonon displacements are related to the microscopic polarization according to
	\be
	p^\alpha_i(\bfR) = -e u^\alpha_i(\bfR),
	\ee
where $\bfR$ labels a lattice site, $\alpha$ is the Cartesian component of the displacement, $i$ labels the basis element within a unit cell, and $e>0$ is the magnitude of the electron charge. The 2D polarization density is related to the microscopic polarization according to 
	\be
	P^\alpha(\bm r, t) = \sum_{\bfR,i} p_i^\alpha(\bfR, t) \delta(\bm r - \bfR - \bfdel_i),
	\ee
where $\bfdel_i$ denotes the $i$-th basis vector within a unit cell.
Consider now a harmonic dependence of the dipole field:
	\be
	p_i^\alpha(\bfR,t) = p_i^\alpha(\bm q,\omega) e^{i[\bm q \cdot(\bfR + \bfdel_i)-\omega t]},
	\label{eq:harmonic_disp}
	\ee
so that 	
	\begin{align}
	P^\alpha(\bm r, \omega) &= \sum_i p_i^\alpha(\bm q, \omega) e^{i\bm q\cdot\bm r}\sum_\bfR \delta(\bm r - \bfR - \bfdel_i) = \sum_i p_i^\alpha(\bm q, \omega) e^{i\bm q\cdot\bm r} \frac{1}{A_c} \sum_\bfG e^{i\bfG\cdot(\bm r - \bfdel_i)},
	\end{align}
where $A_c$ is the unit cell area. The macroscopic (long-wavelength) polarization is the contribution from the $\bfG =0 $ term:
	\be
	P^\alpha(\bm q, \omega) = \frac{1}{A_c}\sum_i p_i^\alpha(\bm q, \omega) \quad \text{(macro)},
	\label{eq:macroP}
	\ee
that is, the macroscopic polarization is just the average of the microscopic polarization within the unit cell.  In Appendix~\ref{appendix:SER}, we analyse the properties within a single unit cell, where the contribution from $\bfG \neq 0$ becomes appreciable.
The 2D current density $j^\alpha$ is related to $ P^\alpha$ via $j^\alpha(\bm q, \omega) = -i\omega P^\alpha(\bm q, \omega)$, implying that the conductivity may be defined through:
    \be
    P^\alpha(\bm q, \omega) = \frac{\sigma_{\alpha\beta}(\bm q,\omega)}{-i\omega} E^\beta(\bm q,\omega),
    \label{eq:P_sigma}
    \ee
with $E^\beta(\bm q,\omega)$ being the electric field. Here and below we also adopt the summation convention over repeated indices, and we do not distinguish between upper and lower indices. 

To relate the conductivity to the phonon Green's function, we consider the response of the system to an \textit{external} electric field $E^\alpha_\text{ext}(\bm r,t)$. The phonon Hamiltonian is modified by the addition of a term corresponding to the interaction energy of a dipole with this external electric field:
	\be
	H_\text{ext} = -\sum_{\bfR , i} p^\alpha_i(\bfR) E_\text{ext}^\alpha(\bfR+\bfdel_i,t) \approx -\sum_{\bfR ,i} p^\alpha_i(\bfR)  E_\text{ext}^\alpha(\bfR,t),
	\ee
where in the approximation we have assumed $E_\text{ext}$ does not vary appreciably within a unit cell. 
Within linear response, the polarization induced by the external field is given by:
	\be
	p^\alpha_i(\bm q, \omega) = \sum_j e^2D^{\alpha\beta}_{ij}(\bm q, \omega) E^\beta_\text{ext}(\bm q, \omega),
    \label{eq:lin_response}
	\ee
where the phonon Green's function is defined as
    \be
    D_{ij}^{\alpha \beta}(\bm q, \omega) = -i\int_0^\infty dt\,   e^{-i\omega t}\, \theta(t) \langle [u^\alpha_i (\bm q,t), u^\beta_j(-\bm q,0)]\rangle.
    \ee
The left-hand side of Eq.~\eqref{eq:lin_response} is related to the macroscopic polarization in the system via Eq.~\eqref{eq:macroP}:
	\be
	P^\alpha(\bm q, \omega) = \frac{e^2}{A_c}\sum_{ij}  D^{\alpha\beta}_{ij}(\bm q, \omega) E_\text{ext}^\beta(\bm q, \omega). 
	\ee
In the regime of linear screening, the \textit{total} and \textit{external} electric fields are related to each other by the dielectric tensor:
\begin{align}
    E_\text{ext}^\alpha(\bm q, \omega) = \varepsilon_{\alpha\beta}(\bm q, \omega) E^\beta(\bm q, \omega),
\end{align}
while the polarization density is related to the electric field by the conductivity, cf. Eq.~\eqref{eq:P_sigma}. This yields
    \be
    \sigma_{\alpha\beta}(\bm q,\omega) = -\frac{ie^2 \omega}{A_c} \sum_{ij} D^{\alpha\gamma}_{ij}(\bm q, \omega) \varepsilon_{\gamma\beta}(\bm q,\omega).
    \ee
The dielectric function is expressed in terms of the conductivity as 
    \be
    \varepsilon_{\alpha\beta}(\bm q, \omega) = \delta_{\alpha\beta} + \frac{2\pi i q}{\omega} \frac{q_\alpha q_\gamma}{q^2} \sigma_{\gamma\beta}(\bm q, \omega),
    \ee
where the projector $q_\alpha q_\gamma/q^2$ describes the screening of longitudinal electric fields. Utilizing this relation, we get:    \be
    \sigma^{-1}_{\alpha\beta}(\bm q, \omega) = -\frac{A_c}{ie^2\omega}\left[ D^{-1}_{\alpha\beta}(\bm q, \omega) - \frac{2\pi e^2}{A_c} \frac{q_\alpha q_\beta}{q} \right],
    \label{eq:sig_D_relation_general}
    \ee
where $D_{\alpha\beta} = \sum_{ij} D^{\alpha\beta}_{ij}$. The result in Eq.~\eqref{eq:sig_D_relation_general} is the general relationship between the optical conductivity and phonon Green's function.   

We now mention some particular important cases. Firstly, for $\bm q=0$, the screening terms vanish and we have the following simple relationship: 
    \be
    \sigma_{\alpha\beta}(\bm q = 0,\omega) = -\frac{ie^2\omega}{A_c} D_{\alpha\beta}(\bm q = 0, \omega),
    \ee
which is a standard result \cite{chitra_wigner_hall,chitra_wigner_long, fukuyama_cdw_magnetic}. Secondly, in a time-reversal symmetric system, where the Hall components vanish, Eq.~\eqref{eq:sig_D_relation_general} reduces to (recall $n = 1/A_c$ for a lattice with a single electron per unit cell):
	\begin{align}
	    \sigma_{\rm T}(q,\omega) = -i ne^2 \omega D_{\rm T}(q,\omega), \qquad
        \frac{\sigma_{\rm L}(q,\omega)}{\varepsilon(q,\omega)} = -i ne^2 \omega D_{\rm L}(q,\omega).
	\end{align}
Finally, in Appendix~\ref{appendix:Magnetic noise} we demonstrate that the magnetic noise tensor in the presence of a perpendicular magnetic field (and hence nonzero $\sigma_{\rm H}$) is expressed in terms of the combination $\tilde \sigma_{\rm T}  = \sigma_{\rm T} + \sigma_{\rm H}^2/\sigma_{\rm L}$, Eq.~\eqref{eqn:noise_BB_ZZ},
which is then related to the phonon Green's function utilizing Eq.~\eqref{eq:sig_D_relation_general}:
    \be
    \tilde \sigma_{\rm T}(\bm q, \omega) = -\frac{ie^2\omega}{A_c} \left[D_{\rm T}(\bm q, \omega) + \frac{D_{\rm H}^2(\bm q, \omega)}{D_{\rm L}(\bm q,\omega)} \right].
    \ee

The analysis in this Appendix is valid for wave vectors $\bm q$ much smaller than the reciprocal lattice vectors. In the following Appendix, we consider the case where large wave vectors become important.

\section{Single-site resolution with local noise spectroscopy}
\label{appendix:SER}

We expect that a qubit probe can be brought to the 2D sample closer than the WC lattice constant, $z_{\rm q} \lesssim a$. In this SSR regime, local electromagnetic noise is strongly sensitive to the in-plane position $\bm r$ of the probe, as fluctuations near an electron site are expected to be enhanced compared to the ones near the middle of a triangle. In this Appendix, we provide the details of our analysis of the magnetic noise in this SSR regime. The opposite regime $z_{\rm q} \gtrsim a$ is considered below in Appendices~\ref{appendix:Magnetic noise} and~\ref{appendix:Electric noise}.

\subsection{Local magnetic noise from a fluctuating WC}

We write the magnetic noise as:
\begin{align}
    {\cal N}^{\rm B}_{\alpha\beta}(\bm r, z_{\rm q},\omega) = \frac{1}{2} \langle \{ B_{\alpha}(\bm r, z_{\rm q},\omega), B_{\beta}(\bm r, z_{\rm q},-\omega) \}\rangle_T,
\end{align}
where we explicitly separated the in-plane position of the probe $\bm r$ from its height $z_{\rm q}$, i.e., $\bm r_{\rm q} = (\bm r, z_{\rm q})$. To simplify the presentation, below we occasionally suppress the explicit dependence on $z_{\rm q}$. The magnetic field $B_{\alpha}(\bm r,\omega)$ at the position of the qubit probe is related  to the charge current inside the sample through the Biot-Savart kernel: 
\begin{align}
    B_{\alpha}(\bm r,\omega) = \int \dd^2\bm r'\, {\cal K}_{\alpha\beta } (\bm r - \bm r') \, j_\beta(\bm r',\omega),
\end{align}
where (in SI units)
\begin{align}
    {\cal K}_{\alpha\beta} (\bm r - \bm r') = - \frac{\mu_0}{4\pi} \frac{\hat{\alpha} \cdot ((\bm r_{\rm q} - \bm r')\times \hat{\beta})}{|\bm r_{\rm q} - \bm r'|^3} \quad \Leftrightarrow \quad
     {\cal K}_{\alpha\beta}(\bm q) = \frac{\mu_0}{2 q} e^{- q z_{\rm q}} \begin{bmatrix}
    0 & q & i q_y\\
    -q & 0 & -iq_x\\
    -i q_y & iq_x & 0 
    \end{bmatrix}.\label{eqn:BS_kernel}
\end{align}
The current density in the monolayer WC is expressed as (see Appendix~\ref{appendix:sigma+D}):
\begin{align}
    j_\alpha(\bm r,\omega) = -i\omega P_\alpha (\bm r, \omega) = -i\omega \sum_{\bfR} p_{\alpha}(\bfR) \delta(\bm r -\bfR) = ie\omega \sum_{\bfR} u_{\alpha}(\bfR) \delta(\bm r -\bfR).
\end{align}
This, together with the fluctuation-dissipation theorem, yields:
\begin{align}
    {\cal N}^{\rm B}_{\alpha\beta}(\bm r, \omega) & \approx 2T n e^2\omega\, \text{Im}\Big[ \sum_{\bfR_1,\bfR_2} {\cal K}_{\alpha\gamma}(\bm r - \bfR_1){\cal K}_{\beta\delta}(\bm r - \bfR_2) D_{\gamma\delta}(\bfR_1 - \bfR_2,\omega) 
    \Big] \notag\\
    & = 2T n e^2\omega\, \text{Im}\Big[ \sum_{\bfG_1,\bfG_2} e^{i\bm r\cdot(\bfG_1 - \bfG_2)} \int_{\rm 1BZ}\frac{\dd^2\bm q}{(2\pi)^2}
    {\cal K}_{\alpha\gamma}(\bm q + \bfG_1){\cal K}_{\beta\delta}(-\bm q - \bfG_2) D_{\gamma\delta}(\bm q,\omega)
    \Big],
    \label{eq:app_nb}
\end{align}
which is Eq.~\eqref{eq:NBr} of the main text.

\subsection{Efficient numerical evaluation of Eq.~\eqref{eq:app_nb} }

The sum over $\bfG_1$ and $\bfG_2$ in Eq.~\eqref{eq:app_nb} converges slowly and is inefficient for numerical evaluation of the noise. To overcome this, we follow the approach outlined in Ref.~\cite{Goldoni:1996} for bilayer WCs and use the Ewald summation technique that turns  Eq.~\eqref{eq:app_nb} into a rapidly convergent sum. We begin by expressing the vector potential through the in-plane currents localized at the 2D WC lattice sites $\bfR$:
    \be
    A_\alpha(\bm r,z) = \frac 1c \sum_{\bfR} \frac{1}{\sqrt{(\bm r -\bfR)^2 + z^2}} j_\alpha(\bfR).
    \ee
Here $z$ is the height above the 2D crystal and we have used the Coulomb gauge $\nabla \cdot \bm A = 0$. This expression is then Fourier transformed
    \be
    A_\alpha(\bm r,z) = \frac 1c\frac{1}{\sqrt N} \sum_{\bm q \in \rm{1BZ}} a(\bm r, z; \bm q) j_\alpha(\bm q),
     \ee
where $N$ is the number of WC lattice sites and 
    \be
    a(\bm r,z;\bm q) \equiv e^{-i\bm q \cdot \bm r} \sum_{\bfR} \frac{e^{i\bm q \cdot (\bm r - \bfR)}}{\sqrt{(\bm r - \bfR)^2 + z^2}}.
    \label{eq:app_a}
    \ee
Efficient evaluation of the sum in Eq.~\eqref{eq:app_a} was worked out in Ref.~\cite{Goldoni:1996} (see Eqs.~(9) and~(13) there), and here we quote the final result:
    \be
    a(\bm r, z; \bm q) = \sqrt n \sum_{\bfG} e^{-i(\bm q + \bfG) \cdot \bm r} \Psi\Big(\frac{(\bm q + \bfG)^2}{4\pi n}, \pi n z^2\Big) + \sqrt n \sum_{\bfR} e^{-i\bm q \cdot \bfR} \Phi\Big(\pi n [(\bm r - \bfR)^2+z^2]\Big),
    \label{eq:app_ewald}
    \ee
where $n$ is the 2D charge density and
    \begin{align}
    \Phi(u) = \sqrt{\frac{\pi}{u}} \erfc(\sqrt u),\qquad
    \Psi(u,v) = \frac 12 \sqrt{\frac{\pi}{u}} \left[e^{\sqrt{4uv}}\erfc(\sqrt u + \sqrt v) + e^{-\sqrt{4uv}}\erfc(\sqrt u - \sqrt v)\right].
    \end{align}
A detailed derivation of these expressions may be found in Appendix A of Ref.~\cite{Goldoni:1996}. 

The magnetic field, given by $\bm B  = \nabla \times \bm A$, may be written as
    \be
    B_\alpha(\bm r, z) = \frac 1c \frac{1}{\sqrt N} \sum_{\bm q \in \rm{1BZ}} b_{\alpha\gamma}(\bm r, z; \bm q) j_\gamma(\bm q), \qquad 
    b_{\alpha\gamma}(\bm r, z; \bm q) \equiv \epsilon_{\alpha\beta\gamma} \partial_\beta a(\bm r,z; \bm q).
    \ee
    Using Eq.~\eqref{eq:app_ewald}, we find for the in-plane components $\alpha = (x,y)$:
    \be
    b_{\alpha\gamma}(\bm r, z; \bm q) = \epsilon_{\alpha\gamma} (2\pi z n^{3/2}) \Big[\sum_{\bfG} e^{-i(\bm q + \bfG) \cdot \bm r} \partial_v\Psi\Big(\frac{(\bm q + \bfG)^2}{4\pi n}, \pi n z^2\Big)+ \sum_{\bfR} e^{-i\bm q \cdot \bfR} \Phi'\Big(\pi n [(\bm r - \bfR)^2+z^2]\Big)\Big],% \quad (\alpha = x,y),
    \ee
where $\epsilon_{\alpha\gamma} \equiv \epsilon_{\alpha z \gamma}$. % and $\partial_v \Psi$ denotes the differentiation with respect to the second argument of the $\Psi$ function. 
For the $\alpha = z$ component, we obtain
    \begin{align}
    b_{z\gamma}(\bm r, z; \bm q) &= \epsilon_{\gamma \beta}\Big[ \sqrt n \sum_{\bfG} [-i(q_\beta + G_\beta)]e^{-i(\bm q + \bfG) \cdot \bm r} \Psi\Big(\frac{(\bm q + \bfG)^2}{4\pi n}, \pi n z^2\Big) \notag\\
    &\quad\quad\quad\quad\quad\quad\quad\quad\quad\quad\quad
    \quad\quad\quad
    + \sqrt n \sum_{\bfR} [2\pi n(r_\beta - R_\beta)]e^{-i\bm q \cdot \bfR} \Phi\Big(\pi n [(\bm r - \bfR)^2+z^2]\Big)\Big].
    \end{align}
The noise is then expressed as 
    \be
    {\cal N}^{\rm B}_{\alpha\beta}(\bm r, z, \omega) = 2T e^2\omega \int_{\rm{1BZ}} \frac{\dd^2\bm q}{(2\pi)^2} ~ b_{\alpha\alpha'}(\bm r,z;\bm q) \, \text{Im}[ D_{\alpha'\beta'}(\bm q, \omega)] \, b_{\beta\beta'}(\bm r, z; -\bm q),
    \ee
which was the form used for calculating the results displayed in Fig.~\ref{fig:SSR} of the main text. There we also used the exact Green's function for a clean WC~\cite{bonsall_elastic_wigner}. This is justified as effects of weak disorder manifest themselves mostly at small wave vectors, while in the SSR regime the noise averages over the entire Brillouin zone.

\subsection{Local $1/T_1$-measurements of the WC}

We now briefly comment on how one accesses the magnetic noise tensor in practice using $1/T_1$ relaxometry. To this end, we first write the Hamiltonian that governs the dynamics of the local qubit probe:
\begin{align}
    \hat{H}_{\rm q} = \frac{\hbar \omega_{\rm q}}{2} \hat{\bm \sigma}\cdot \hat{n}_{\rm q} + \frac{g \mu_B}{2}  \hat{\bm \sigma}\cdot \hat{\bm B}(\bm r_{\rm q},t),
\end{align}
where $g$ is the $g$-factor of the probe and $\hbar \omega_{\rm q}$ is its splitting along the quantization axis $\hat{n}_{\rm q}$.  In $1/T_1$-experiments, one studies the decay rate of the qubit initially polarized along $\hat{n}_{\rm q}$. Using Fermi's golden rule, one finds~\cite{Langsjoen:2012}:
\begin{align}
    \frac{1}{T_1}  = \left(\frac{g\mu_B}{2}\right)^2 {\cal N}^{\rm B}_{-+}(\omega_{\rm q}),~\label{eqn:1/T_1_gen}
\end{align}
where $\hat{B}^{\pm} = \hat{B}_{x'} \pm i\hat{B}_{y'}$ (here $(x',y',z' = \hat{n}_{\rm q})$ form a mutually orthogonal triad). Strictly speaking, Eq.~\eqref{eqn:1/T_1_gen} is correct only for time-reversal symmetric situations; if this symmetry is broken (spontaneously or via an applied magnetic field), Eq.~\eqref{eqn:1/T_1_gen} is valid only up to the leading order in  $(\omega_{\rm q} z_{\rm q}/c)^2 \ll 1$, assumed throughout the paper -- see also Appendix~\ref{appendix:Magnetic noise}. In Fig.~\ref{fig:SSR} of the main text, where we discuss the SSR regime, we primarily consider the quantization axis $\hat {n}_{\rm q}$ to be aligned along $\hat{z}$.

\subsection{Local $1/T_2$-measurements of the WC}

In $1/T_2$-experiments, one performs an analog of a spin-echo or Ramsey pulse sequence on the qubit~\cite{abragam1961principles}. During such a pulse sequence, the qubit initially aligned in the plane perpendicular to $\hat{n}_{\rm q}$ precesses under the magnetic field component along $\hat{n}_{\rm q}$. For a noisy magnetic field, on average the qubit will display dephasing. Similarly to the $1/T_1$-rate, the dephasing rate can be related to the magnetic noise coming from the sample~\cite{machado2022quantum}. Specifically, this rate is encoded in the time decay of  ${\cal P}(\tau) = \exp(-2\langle\varphi^2(\tau)\rangle$), where
\begin{align}
    \langle\varphi^2(\tau)\rangle = \int_{-\infty}^\infty    \frac{d\omega}{2\pi}  W_{\tau}(\omega) \, \hat{n}_{{\rm q},\alpha} {\cal N}^{\rm B}_{\alpha \beta}(\omega)\hat{n}_{{\rm q},\beta}.
\end{align}
Here $W_{\tau}(\omega)$ is the filtering function, which encodes the pulse-sequence used in the experiment. For the traditional spin-echo sequence, we have~\cite{machado2022quantum}
\begin{align}
    W_{\tau}(\omega)  = (2g\mu_B)^2 \frac{\sin^4(\tau \omega/4)}{\omega^2}.
\end{align}
If the noise is frequency independent, such as in the liquid state, we get:
\begin{align}
    \Big[\frac{1}{T_2}\Big]_\mathrm{Liq} = (g\mu_B)^2 \hat{n}_{{\rm q},\alpha} [{\cal N}^{\rm B}_{\alpha \beta}]_\mathrm{Liq}\hat{n}_{{\rm q},\beta}.
\end{align}
Crudely, the $1/T_2$-rate is expected to be of the same order as the $1/T_1$-rate, Eq.~\eqref{eqn:1/T_1_gen}. The crucial difference between the two manifests when the noise depends on frequency, such as in the WC state. In this case, the $1/T_1$-rate is determined by the qubit splitting $\omega_{\rm q}$ (for NVs, $\omega_{\rm q} \simeq 2.7\,$GHz), whereas the $1/T_2$-rate essentially senses the magnetic noise at frequencies related to the pulse sequence (typically in the sub-MHz range).

\subsection{Estimates of the magnetic noise in the liquid phase for TMDs}

The feasibility of both types of experiments, $1/T_1$ relaxometery and $1/T_2$ spin-echo-like measurements, is directly encoded in the magnetic noise strength.

For the reference magnetic noise, we use that of the liquid state, which we write in the following convenient form:
    \be
    [{\cal N}_{zz}^{\rm B}]_\mathrm{Liq} \approx 3.4\, \text{pT}^2 \times \text{Hz}^{-1} \times \left(\frac{\sigma_0}{\sigma_Q}\right) \left(\frac{T}{10 \, \text K}\right) \left(\frac{10 \, \text{nm}}{z_{\rm q}}\right)^2,\label{eqn:noise_est}
    \ee
where $\sigma_Q = 2e^2/h$ is the quantum of conductance, and we have taken $T = 10\, \text K$ and $z_{\rm q} = 10\,\text{nm}$ as the reference temperature and probe height, respectively. The estimate of the magnetic noise reduces to estimating $\sigma_0$, which is related to the mobility $\mu$ as:
\begin{align}
    \frac{\sigma_0}{\sigma_Q} = \frac{e n \mu}{\sigma_Q} \approx 1.5 \times \Big(\frac{n}{5\times 10^{11} \, \text{cm}^{-2}}\Big) \Big(\frac{\mu}{1500 \, \text{cm}^{2}/(\text{V}\, \text{s})}\Big),
\end{align}
where we have used $1500 \, \text{cm}^{2}/(\text{V}\, \text{s})$ as the reference mobility -- this value is close to the one reported in Ref.~\cite{Larentis:2018} for monolayer and bilayer MoSe$_2$. The relevant electron densities in monolayer TMD samples are $n \simeq 5\times 10^{11} \, \text{cm}^{-2}$. For monolayer TMDs, we, therefore, get $[{\cal N}_{zz}^{\rm B}]_\mathrm{Liq} \simeq 5\, \text{pT}^2 \times \text{Hz}^{-1}$, which is within the sensitivity of current qubit sensors~\cite{andersen2019electron}. Let us remark that  the TMD value $\sigma_0/\sigma_Q\simeq 1$ appears to be rather small, i.e., the TMD samples are rather dirty. Indeed, for more traditional semiconductors that also exhibit signatures of the WC phase, this ratio can be several orders of magnitude larger~\cite{spivak2010colloquium}. We anticipate that fabrication of cleaner TMD samples will emerge in the foreseeable future, which will then increase the feasibility of our proposal.

In bilayer TMD samples, there are two aspects that make the magnetic noise much stronger~\cite{Zhou_2021}: i) bilayer WCs in TMDs are stable up to anomalously high electron densities (about an order of magnitude larger than the typical monolayer density where putative WCs are present) so that one can easily get $[{\cal N}_{zz}^{\rm B}]_\mathrm{Liq} \simeq 50\, \text{pT}^2 \times \text{Hz}^{-1}$; and ii) further enhancement can come from the fact that the melting temperature in bilayer WCs is $T_c \approx 40 \,$K. From Eq.~\eqref{eqn:noise_est} it then follows that, even without taking into account the proliferation of fluctuations near the melting transition, the noise can be further enhanced by another factor of four $[{\cal N}_{zz}^{\rm B}]_\mathrm{Liq} (40\,
\text{K}) \simeq 200\, \text{pT}^2 \times \text{Hz}^{-1}$. We also remark that if the qubit-sample distance $z_{\rm q}$ is larger than the interlayer separation, the noise from a bilayer will be amplified by roughly a factor of two since the signal comes from two (as opposed to one) layers.

So far we have estimated the magnetic noise in the liquid state. Figure~\ref{fig:freq_scaling} of the main text shows that the (spatially averaged) noise from the WC state is expected to be comparable to $[{\cal N}_{zz}^{\rm B}]_\mathrm{Liq}$ so long as $\omega$ is close to the pinning resonance $\omega_0$. For $\omega \ll \omega_0$, we expect a suppression by $(\omega/\omega_0)^2$. In the SSR, on the other hand, one can easily get an order of magnitude enhancement near an electron site -- see Fig.~\ref{fig:SSR}.

\section{Magnetic noise tensor from two-dimensional systems with nonzero Hall conductivity}
\label{appendix:Magnetic noise}

In this Appendix, we relate the magnetic noise tensor in Eq.~\eqref{eq:NB} to the electromagnetic correlation functions of the two-dimensional sample. Our analysis closely follows that of Ref.~\cite{Agarwal2017Magnetic}, except here we consider the case, where time-reversal symmetry can be (spontaneously or via applied magnetic field) broken, which leads, in particular, to the development of the Hall conductivity $\sigma_{\rm H}(\bm q,\omega)$. We assume that the system is translationally- and rotationally-invariant (for Wigner crystals, this assumption approximately holds only for the effective description in terms of low-energy low-momenta phonons, i.e., for $z_{\rm q}\gtrsim a$), which allows one to write the conductivity tensor as~\cite{PhysRevMaterials.2.014003}:
\begin{align}
    \sigma_{\alpha\beta} (\bm q, \omega) = \frac{q_\alpha q_\beta}{q^2}\sigma_{\rm L}(q,\omega) + \Big(\delta_{\alpha\beta} - \frac{q_\alpha q_\beta}{q^2}\Big)\sigma_{\rm T}(q,\omega) + \varepsilon_{\alpha\beta} \sigma_{\rm H}(q,\omega),  \label{eqn:sigma_tensor}
\end{align}
where $\varepsilon_{\alpha\beta}$ is the two-dimensional anti-symmetric Levi-Civita tensor ($\varepsilon_{xy} = - \varepsilon_{yx} = 1$). The form in Eq.~\eqref{eqn:sigma_tensor} imposes the magnetic noise tensor to acquire the following structure:
\begin{align}
    {\cal N}^{\rm B}_{\alpha\beta}(\omega) = \begin{bmatrix}[1.2]
    {\cal N}^{\rm B}_{xx}(\omega) & {\cal N}^{\rm B}_{xy}(\omega) & 0\\
    -{\cal N}^{\rm B}_{xy}(\omega) & {\cal N}^{\rm B}_{xx}(\omega) & 0\\
    0 & 0 & {\cal N}^{\rm B}_{zz}(\omega) 
    \end{bmatrix}, \label{eqn:N_BB}
\end{align}
i.e., there are three independent noise components (${\cal N}^{\rm B}_{xx}$, ${\cal N}^{\rm B}_{xy}$, and ${\cal N}^{\rm B}_{zz}$), which we turn to compute below.

When evaluating the magnetic noise tensor, we employ the fluctuation-dissipation theorem which relates this tensor to the respective response function (throughout the text, we set $k_B = 1$):
\begin{align}
    {\cal N}^{\rm B}_{\alpha\beta}(\omega) =  \hbar \coth\Big(\frac{\hbar\omega}{2 T} \Big) \text{Im}[\chi^{\rm B}_{\alpha\beta}(\omega)],
\end{align}
where
\begin{align}
    \chi^{\rm B}_{\alpha\beta}(\omega) \equiv \int dt \, e^{i\omega t} \chi^{\rm B}_{\alpha\beta}(t) \text{ and } \chi^{\rm B}_{\alpha\beta}(t - t') \equiv  -i\theta(t- t') \langle [B_{\alpha}(\bm r_{\rm q},t),B_{\beta}(\bm r_{\rm q},t')] \rangle_T.
\end{align}
A possible way to compute $\chi^{\rm B}_{\alpha\beta}(\omega)$ is to place a magnetic dipole moment at the location of the probe, $\bm m_0\delta(\bm r)\delta(z-z_{\rm q}) e^{-i\omega t}$, and then evaluate the induced magnetic field $\bm B(\bm r_{\rm q}, \omega)$. For future reference, we write $\delta(\bm r) = \displaystyle \int \frac{d^2 \bm q}{(2\pi)^2} e^{i\bm q \cdot \bm r}$, which, instead of a point-like magnetic dipole moment, allows us to consider a two-dimensional sheet with magnetization profile of the form $\bm m_0 e^{i\bm q \cdot \bm r} \delta(z-z_{\rm q}) e^{-i\omega t}$ -- this representation is particularly useful as one can now employ in-plane translational invariance, where the in-plane momentum $\bm q$ is a good conserving number.

\subsection{ Evaluation of ${\cal N}^{\rm B}_{zz}(\omega)$
}

We begin by evaluating ${\cal N}^{\rm B}_{zz}(\omega)$. To do so, one aligns the magnetic dipole moment along the $z$-axis $\bm m_0 = m_0 \hat{z}$. In this case, it is convenient to think about the magnetization sheet as if it gives rise to an external current density of the form $\bm J_{\rm ext} = \bm j_{\rm ext} \delta(z-z_{\rm q})e^{i\bm q \cdot \bm r-i\omega t}$, where $ \bm j_{\rm ext} = i q m_0 \hat{q}\times\hat{z}$. This current density, in turn, enters Maxwell's equations as a source term:
\begin{align}
    \nabla \times \bm B = \frac{1}{c^2} \frac{\partial \bm E}{\partial t} + \mu_0(\bm j \delta(z) + \bm J_{\rm ext}). \label{eqn:Max_v0}
\end{align}
Here $\bm j$ is the two-dimensional current density flowing in the sample, and it can develop in response to the drive $\bm J_{\rm ext}$. Our task at hand is to evaluate the magnetic field $\bm B$ at $z = z_{\rm q}$, which we do by solving the Maxwell equations in each of the three regions $z < 0$, $0 < z < z_{\rm q}$, and $z_{\rm q} < z$, and then match the solutions using the Fresnel boundary conditions at $z = 0$ and $z = z_{\rm q}$:
\begin{align}
    E_z^+ - E_z^- = \frac{\rho}{\varepsilon_0},\quad B_z^+ = B_z^-,\quad \hat{z}\times (\bm B^+ - \bm B^-) = \mu_0 \bm j,\quad \bm E_t^+ = \bm E_t^-. \label{eqn:bc_v0}
\end{align}

We turn to revisit Maxwell's equations, as below we decompose vectors as:
\begin{align}
    \bm E(z,\bm q,\omega) = E_\parallel(z,\bm q,\omega) \hat{q} + E_\perp(z,\bm q,\omega) \hat{q}\times\hat{z} + E_z(z,\bm q,\omega) \hat{z}.
\end{align}
For future reference, we note that a simple vector analysis gives:
\begin{align}
    \nabla \times (E_\parallel \hat{q}) = -\partial_z E_\parallel \, \hat{q}\times\hat{z},\quad  
    \nabla \times (E_\perp \hat{q}\times\hat{z}) = \partial_z E_\perp\hat{q} - iqE_\perp\hat{z},\quad
    \nabla \times (E_z \hat{z}) = i q E_z\hat{q}\times\hat{z}.
\end{align}
Substituting this into the Faraday law, we get:
\begin{align}
    B_\parallel = \frac{\partial_z E_\perp}{i\omega},\quad
    B_\perp = \frac{1}{i\omega}( iq E_z - \partial_z E_\parallel),\quad
    B_z = -\frac{iq E_\perp}{i\omega}.
\end{align}
The remaining Maxwell equations then read:
\begin{gather}
    i q E_\parallel + \partial_z E_z = \frac{\rho}{\varepsilon_0} \delta(z),\\
    \partial_z B_\perp = -\frac{i\omega}{c^2} E_\parallel + \mu_0 j_\parallel \delta(z),\\
    iq B_z - \partial_z B_\parallel = -\frac{i\omega}{c^2} E_\perp + \mu_0 (j_\perp \delta(z) + iqm_0\delta(z-z_{\rm q})),\\
    -i q B_\perp = -\frac{i\omega}{c^2} E_z,
\end{gather}
where $\rho$ is the two-dimensional charge density related to the current density via the continuity equation.
We further have $j_\parallel = \sigma_{\rm L} E_\parallel - \sigma_{\rm H} E_\perp$ and $j_\perp = \sigma_{\rm T} E_\perp + \sigma_{\rm H} E_\parallel$ -- we note that the Hall conductivity mixes the longitudinal and transverse sectors. Finally, we point out that the boundary condition involving the current density modifies to:
\begin{align}
    B^+_\perp - B^-_\perp = \mu_0 j_\parallel,\quad  B^+_\parallel - B^-_\parallel = -\mu_0 j_\perp.
\end{align}

With the above decomposition, the solution of Maxwell's equations can be written as (the magnetization sheet can only emit  radiation away):
\begin{align}
E_{\parallel(\perp)}(z) = 
\begin{cases}
E_{1,\parallel(\perp)} e^{iq_z (z - z_{\rm q})} & z_{\rm q} < z\\
\alpha_{\parallel(\perp)} e^{-iq_z (z - z_{\rm q})} + \beta_{\parallel(\perp)} e^{i q_z z} & 0 < z < z_{\rm q}\\
E_{2,\parallel(\perp)} e^{-iq_z z} & z < 0
\end{cases},\, 
E_z(z) = 
\displaystyle \displaystyle \frac{q}{q_z} \begin{cases} \displaystyle -
E_{1,\parallel} e^{iq_z (z - z_{\rm q})} & z_{\rm q} < z\\
 \alpha_\parallel e^{-iq_z (z - z_{\rm q})} - \beta_\parallel e^{i q_z z} & 0 < z < z_{\rm q}\\
E_{2,\parallel} e^{-iq_z z} & z < 0
\end{cases}. \label{eqn:E_z_ansatz}
\end{align}
Here $E_{1,\parallel(\perp)}$ ($E_{2,\parallel(\perp)}$) represents the in-plane electric field at $z = z_{\rm q}$ ($z = 0$). We have also defined:
\begin{align}
q_z \equiv \begin{cases}
\sqrt{\omega^2/c^2 - q^2} & \omega \geq qc\\
i\sqrt{q^2 - \omega^2/c^2 } & \omega < qc
\end{cases}.
\end{align}
The coefficients $\alpha_\parallel, \beta_\parallel$ and $\alpha_\perp, \beta_\perp$
are found through the continuity of the tangential component of the electric field:
\begin{gather}
    \alpha_{\parallel(\perp)} = \frac{E_{2,\parallel(\perp)} - E_{1,\parallel(\perp)} e^{-iq_z z_{\rm q}} }{e^{iq_z z_{\rm q}} - e^{-iq_z z_{\rm q}}},\quad
    \beta_{\parallel(\perp)} = \frac{E_{1,\parallel(\perp)} - E_{2,\parallel(\perp)} e^{-iq_z z_{\rm q}} }{e^{iq_z z_{\rm q}} - e^{-iq_z z_{\rm q}}}.
\end{gather}
The boundary conditions at $z = 0$ give:
\begin{align}
   \beta_\parallel  = -\frac{q_z c^2 }{2 \omega} \mu_0 j_\parallel,\quad \beta_\perp = -\frac{\omega}{2 q_z} \mu_0 j_\perp.
\end{align}
The boundary conditions at $z = z_{\rm q}$ give:
\begin{align}
    \alpha_\parallel = 0,\quad   \alpha_\perp = -\frac{i \mu_0 m_0 q \omega}{2 q_z}.
\end{align}
Collecting all of the above results, we evaluate the magnetic field at the location of the qubit:
\begin{align}
    \bm B(\bm r_{\rm q},\omega) = \hat{z} \int \frac{d^2 \bm q}{(2\pi)^2} \frac{i \mu_0 m_0 q^2}{2 q_z}\Big[ 1 - \frac{e^{2i q_z z_{\rm q}}}{1 + 2q_z/(\mu_0 \omega \tilde{\sigma}_{\rm T})}\Big],\quad
    \tilde{\sigma}_{\rm T} = \sigma_{\rm T} + \frac{\sigma_{\rm H}^2}{\sigma_{\rm L}} \Big[ 1 + \frac{2\omega \varepsilon_0}{q_z\sigma_{\rm L}} \Big]^{-1}. 
\end{align}
From this, we obtain ${\cal N}_{zz}^{\rm BB}(\omega)$:
\begin{align}
    {\cal N}_{zz}^{\rm B}(\omega) \approx \frac{T \mu_0}{2\pi\omega} \int_0^\infty dq \, q^2\, e^{-2qz_{\rm q}} \, \text{Im}\Big[ \frac{-1}{1 + 2iq/(\mu_0 \omega \tilde{\sigma}_{\rm T}(q,\omega))} \Big] \approx \frac{T \mu_0^2 }{16\pi z_{\rm q}^2} \int_0^\infty dx\, x\, e^{-x}\,  \text{Re}\Big[ \tilde{\sigma}_{\rm T}\Big(\frac{x}{2z_{\rm q}}, \omega \Big) \Big],  \label{eqn:noise_BB_ZZ}
\end{align}
where in the first identity we approximated: i) $\hbar\coth(\hbar\omega/2T)\approx 2T/\omega$ (this substitution is often referred to as the classical limit) since typically $T$ is much larger than the relevant energy $\hbar \omega$; and ii) $q_z\approx i q$, i.e., we neglected the contribution from propagating waves, as their phase space is negligible compared to that of the evanescent waves. In the second identity, we assumed that the frequency $\omega$ is small or, more rigorously, we expanded to the leading order in the small parameter $\omega z_{\rm q}/c$. We note that the Hall conductivity manifests only through the substitution $\sigma_{\rm T} \to \tilde{\sigma}_{\rm T}$ -- other than that, Eq.~\eqref{eqn:noise_BB_ZZ} reproduces the result of Ref.~\cite{Agarwal2017Magnetic} derived for $\sigma_{\rm H} = 0$. Within the same approximations as in Eq.~\eqref{eqn:noise_BB_ZZ}, we further have:
\be
    \tilde{\sigma}_{\rm T} \approx \sigma_{\rm T} + \frac{\sigma_{\rm H}^2}{\sigma_{\rm L}}.
\ee

\subsection{Evaluation of the in-plane components of the magnetic tensor}

To evaluate ${\cal N}^{\rm B}_{xx}(\omega)$ and ${\cal N}^{\rm B}_{xy}(\omega)$, one now aligns $\bm m_0$ along the $xy$-plane, which allows us to write $\bm m_0 = m_\parallel\hat{q} + m_\perp \hat{q}\times\hat{z}$ -- below, we consider these two contributions separately. For concreteness, we shall assume that $\bm m_0 = m_0\hat{x}$ so that $m_\parallel = m_0\cos\vartheta$ and $m_\perp = m_0\sin\vartheta$, where $\vartheta$ is the polar angle of $\bm q$. All the analysis above, in particular the free-space solution in Eq.~\eqref{eqn:E_z_ansatz}, is applicable here as well, except for the boundary conditions at $z = z_{\rm q}$, which we turn to derive below.

\emph{Fresnel boundary conditions for $m_\perp$}---We note that the magnetization $\bm M = \bm m_0\delta(z-z_{\rm q}) e^{i\bm q\cdot\bm r-i\omega t}$ is such that $\nabla\times \bm M$ has a nonzero out-of-plane component, which makes the above picture of external currents no longer intuitive, cf. Eq.~\eqref{eqn:Max_v0}. Instead, we now will work with the free-field $\bm H$ ($\bm B = \mu_0(\bm H + \bm M)$) and write the Maxwell equations near $z = z_{\rm q}$ as:
\begin{align}
     \nabla \cdot \bm E = 0, \quad \nabla \cdot (\bm H + \bm M) = 0, \quad \nabla \times \bm E = -\mu_0\frac{\partial}{\partial t}(\bm H + \bm M), \quad \nabla \times \bm H = \frac{\varepsilon_0}{c}\frac{\partial\bm E}{\partial t}. \label{eqn:Maxwell_H-field}
\end{align}
 From these equations, we get the boundary conditions at $z = z_{\rm q}$:
\begin{align}
    E_z^+ = E_z^-,\quad B_z^+ = B_z^-,\quad E_\parallel^+ - E_\parallel^- = - i\omega\mu_0 m_\perp,\quad E_\perp^+ = E_\perp^-  ,\quad \bm B_t^+ = \bm B_t^-.
\end{align}
Given these equations, we obtain a few useful relations that fix the parameters entering the ansatz in Eq.~\eqref{eqn:E_z_ansatz} (note that the tangential component of the electric field is no longer continuous at $z = z_{\rm q}$):
\begin{align}
    E_{1,\perp} = \alpha_\perp + \beta_\perp e^{iq_z z_{\rm q}},\quad 
    E_{2,\perp} = \alpha_\perp e^{iq_z z_{\rm q}} + \beta_\perp,\quad 
    E_{1,\parallel} = \alpha_\parallel + \beta_\parallel e^{iq_z z_{\rm q}} - i\omega\mu_0m_\perp,\quad 
    E_{2,\parallel} = \alpha_\parallel e^{iq_z z_{\rm q}} + \beta_\parallel. \label{eqn:E_field_perp_ansatz}
\end{align}
Furthermore, the boundary conditions at $z = 0$ and $z = z_{\rm q}$ give:
\begin{align}
   \beta_\parallel  = -\frac{q_z ac^2 }{2 \omega} \mu_0 j_\parallel,\quad \beta_\perp = -\frac{\omega}{2 q_z} \mu_0 j_\perp,\quad \alpha_\parallel = \frac{1}{2}i\omega \mu_0 m_\perp,\quad   \alpha_\perp = 0.
\end{align}
Solving these equations, we obtain a few useful relations:
\begin{align}
    \beta_\perp = -\frac{\sigma_{\rm H}}{\sigma_{\rm T}} \frac{E_{2,\parallel}}{1 + 2 q_z/(\mu_0\omega\sigma_{\rm T})},\quad
    \beta_\parallel = -\frac{\alpha_\parallel e^{iq_z z_{\rm q}}}{1 + 2\omega \varepsilon_0/(q_z\tilde{\sigma}_{\rm L})},\quad \tilde{\sigma}_{\rm L} = \sigma_{\rm L} + \frac{\sigma^2_{\rm H}}{\sigma_{\rm T}} \Big[ 1 + \frac{2q_z}{\mu_0 \omega \sigma_{\rm T}} \Big]^{-1}.
\end{align}

\emph{Fresnel boundary conditions for $m_\parallel$}---From Eq.~\eqref{eqn:Maxwell_H-field}, we  obtain the boundary conditions at $z = z_{\rm q}$:
\begin{align}
     E_z^+ = E_z^-,\quad B_z^+ - B_z^- = - iq\mu_0 m_\parallel,\quad E_\perp^+ - E_\perp^- =  i\omega\mu_0 m_\parallel,\quad E_\parallel^+ = E_\parallel^-  ,\quad \bm B_t^+ = \bm B_t^-.
\end{align}
Eq.~\eqref{eqn:E_field_perp_ansatz} then modifies to:
\begin{align}
    E_{1,\perp} = \alpha_\perp + \beta_\perp e^{iq_zz_{\rm q}} +  i\omega\mu_0m_\parallel,\quad
    E_{2,\perp} = \alpha_\perp e^{iq_zz_{\rm q}} + \beta_\perp,\quad
    E_{1,\parallel} = \alpha_\parallel + \beta_\parallel e^{iq_zz_{\rm q}},\quad
    E_{2,\parallel} = \alpha_\parallel e^{iq_zz_{\rm q}} + \beta_\parallel.
\end{align}
The boundary conditions at $z = 0$ and $z = z_{\rm q}$ give:
\begin{align}
   \beta_\parallel  = -\frac{q_z c^2 }{2 \omega} \mu_0 j_\parallel,\quad \beta_\perp = -\frac{\omega}{2 q_z} \mu_0 j_\perp,\quad
   \alpha_\parallel = 0,\quad   \alpha_\perp = -\frac{1}{2}i\omega \mu_0 m_\parallel.
\end{align}
Solving the boundary conditions, we further obtain a few additional useful relations:
\begin{align}
    \beta_\parallel = \frac{\sigma_{\rm H}}{\sigma_{\rm L}}\frac{E_{2,\perp}}{1 + 2\omega \varepsilon_0/(q_z{\sigma}_{\rm L})},\quad
    \beta_\perp = -\frac{\alpha_\perp e^{iq_zz_{\rm q}}}{1 + 2q_z /(\mu_0 \omega \tilde{\sigma}_{\rm T})}.
\end{align}

\emph{Evaluation of the in-plane magnetic noise tensor}---Collecting all the above results, we compute the magnetic field at the location of the probe at $z= z_{\rm q}$:
\begin{gather}
    B_x(\bm r_{\rm q},\omega) = \int_0^\infty \frac{q dq }{2\pi}  \frac{i\mu_0 m_0 \omega^2}{4q_zc^2}  \Big[ 1 + \frac{e^{2iq_zz_{\rm q}}}{1 + 2\omega \varepsilon_0/(q_z{\tilde{\sigma}}_{\rm L})} \Big]
    +  \int_0^\infty \frac{q dq }{2\pi} \frac{i\mu_0 m_0 q_z}{4} \Big[ 1 + \frac{e^{2i q_z z_{\rm q}}}{1 + 2q_z/(\mu_0 \omega \tilde{\sigma}_{\rm T})}\Big]
\end{gather}
and
\begin{align}
     B_y(\bm r_{\rm q},\omega)  = & - \int_0^\infty \frac{q dq }{2\pi} \frac{\sigma_{\rm H}}{\sigma_{\rm T}} \frac{i\mu_0 m_0 q_z}{4} \frac{e^{2iq_zz_{\rm q}}}{1 + 2q_z/(\mu_0 \omega \sigma_{\rm T }) } \Big[ 1 - \frac{1}{1 + 2\omega\varepsilon_0/(q_z \tilde{\sigma}_{\rm L})}\Big] \notag\\
     &\qquad\qquad\qquad\qquad\qquad
     - \int_0^\infty \frac{q dq }{2\pi} \frac{\sigma_{\rm H}}{\sigma_{\rm L}} \frac{i\mu_0 m_0 \omega^2}{4q_zc^2} \frac{e^{2iq_zz_{\rm q}}}{1 + 2\omega \varepsilon_0/(q_z{\sigma}_{\rm L})} \Big[ 1 - \frac{1}{1 + 2q_z/(\mu_0 \omega \tilde{\sigma}_{\rm T})}\Big].
\end{align}
From these expressions and  within the same approximations as in Eq.~\eqref{eqn:noise_BB_ZZ}, we get the remaining components of the magnetic noise tensor:
\begin{align}
    {\cal N}^{\rm B}_{xx}(\omega) & \approx \int_0^\infty dq  \frac{T \mu_0  \omega }{4\pi c^2} e^{-2qz_{\rm q}}\, \text{Im}\Big[ \frac{1}{1 + 2\omega \varepsilon_0/(iq{\tilde{\sigma}}_{\rm L})}\Big] 
    + \int_0^\infty dq \frac{T \mu_0  q^2 }{4 \pi \omega} e^{-2 q z_{\rm q}}\, \text{Im}\Big[ \frac{-1}{1 + 2iq/(\mu_0 \omega \tilde{\sigma}_{\rm T})} \Big] \notag\\
    & \approx \frac{ T \mu_0^2   }{32 \pi z_{\rm q}^2}
    \int_0^\infty dx \,  x \, e^{-x}\,   \text{Re}\Big[ \tilde{\sigma}_{\rm T}\Big(\frac{x}{2z_{\rm q}}, \omega \Big) \Big] \approx \frac{1}{2}{\cal N}^{\rm B}_{zz}(\omega),\\
    {\cal N}^{\rm B}_{xy}(\omega) & \approx 0.
\end{align}
We, therefore, conclude that to the leading order in $\omega z_{\rm q}/c$, the off-diagonal components of the magnetic noise tensor vanish.

\section{Electric noise tensor from two-dimensional systems with nonzero Hall conductivity}
\label{appendix:Electric noise}

In the main text, we have primarily focused on the magnetic noise sensing of WCs. However, electrical noise sensing is also possible via both electric noise qubits and SNOM detectors.
For completeness, in this Appendix we therefore evaluate the electric noise tensor following step-by-step the preceding Appendix~\ref{appendix:Magnetic noise}. Much of the discussion above is applicable here as well, including the structure of the noise tensor in Eq.~\eqref{eqn:N_BB} and the free-space solution in Eq.~\eqref{eqn:E_z_ansatz}.
The primary difference compared to the analysis above is that instead of a magnetic dipole moment placed at the location of the qubit, we now consider an electric dipole moment and evaluate the induced electric field. In particular, instead of a magnetization vector $\bm M = \bm m_0\delta(z-z_{\rm q}) e^{i\bm q\cdot\bm r-i\omega t}$, we will work with a polarization vector $\bm P = \bm p_0\delta(z-z_{\rm q}) e^{i\bm q\cdot\bm r-i\omega t}$. This difference between the two calculations manifests only trough the boundary conditions at $z = z_{\rm q}$.

\subsection{Evaluation of the in-plane components of the electric tensor}

In case $\bm p_0$ is aligned within the $xy$-plane (so that one can write $\bm p_0 = p_\parallel \hat{q} + p_\perp \hat{q}\times\hat{z}$), the polarization vector enters the Maxwell equations through an external current density $\bm J_{\rm ext} = \partial_t \bm P$, cf. Eq~\eqref{eqn:Max_v0}. This implies that the boundary conditions, at both $z =0$ and $z = z_{\rm q}$, are given by Eq.~\eqref{eqn:bc_v0}. Following similar algebra as above, we evaluate the electric field at $z = z_{\rm q}$:
\begin{gather}
    E_{\parallel}(z_{\rm q},\bm q,\omega) =
    \frac{i \mu_0 p_\parallel c^2 q_z}{2} \Big[ 1 -  \frac{e^{2iq_zz_{\rm q}}}{ 1 +  2 \omega\varepsilon_0/(q_z\tilde{\sigma}_{\rm L})} \Big] + \frac{\sigma_{\rm H}}{\sigma_{\rm L}} \frac{i \mu_0 p_\perp \omega^2}{2q_z} \frac{1}{1 + 2\omega\varepsilon_0/(q_z\sigma_{\rm L})} \frac{e^{2iq_zz_{\rm q}}}{ 1 + \mu_0 \omega \tilde{\sigma}_{\rm T}/(2 q_z)},
    \\
    E_{\perp}(z_{\rm q},\bm q,\omega) = - \frac{\sigma_{\rm H}}{\sigma_{\rm L}} \frac{i \mu_0 p_\parallel c^2 q_z}{2} \frac{1}{1 + 2q_z/(\mu_0\omega\sigma_{\rm T})} \frac{e^{2iq_zz_{\rm q}}}{ 1 + q_z \tilde{\sigma}_{\rm L}/(2 \omega\varepsilon_0)}
    +
    \frac{i \mu_0 p_\perp \omega^2}{2q_z} \Big[ 1 -  \frac{e^{2iq_zz_{\rm q}}}{ 1 +  2q_z/(\mu_0\omega \tilde{\sigma}_{\rm T})} \Big].
\end{gather}
From these expressions and within the same approximations as in Eq.~\eqref{eqn:noise_BB_ZZ}, we get:
\begin{align}
    {\cal N}^{\rm E}_{xx}(\omega) & \approx \int_0^\infty dq \frac{T\mu_0 c^2 q^2}{4\pi \omega}e^{-2qz_{\rm q}}\, \text{Im}\Big[ \frac{1}{1 + 2\omega\varepsilon_0/(i q \tilde{\sigma}_{\rm L})}
    \Big] + \int_0^\infty dq \frac{T\mu_0 \omega}{4\pi}e^{-2qz_{\rm q}}\,
    \text{Im}\Big[ 
    \frac{-1}{ 1 +  i q/(\mu_0\omega \tilde{\sigma}_{\rm T})}
    \Big]\\
    & \approx \int_0^\infty dq \frac{T\mu_0 c^2 q^3}{8\pi \omega^2\varepsilon_0 }e^{-2qz_{\rm q}}\, \text{Re}\Big[ \frac{\tilde{\sigma}_{\rm L}(q,\omega)}{\varepsilon(q,\omega) }
    \Big] =  \frac{T}{128\pi\omega^2\varepsilon_0^2 z_{\rm q}^4} \int_0^\infty dx \, x^3 \, e^{-x} \, \text{Re}\Big[ \frac{\tilde{\sigma}_{\rm L}(x/(2z_{\rm q}),\omega)}{\varepsilon(x/(2z_{\rm q}),\omega) } \Big],\\
    {\cal N}^{\rm E}_{xy}(\omega) & \approx 0.
\end{align}
where $\varepsilon(q,\omega) = 1 + iq\tilde{\sigma}_{\rm L}/(2\omega\varepsilon_0)$ is the permittivity that captures Coulomb screening effects. As it was for the magnetic noise tensor, here we also find that the off-diagonal components of the electric noise tensor are suppressed in the leading order in $\omega z_{\rm q}/c$. To this same order of approximation, we further have $\tilde \sigma_{\rm L} = \sigma_{\rm L}$.

\subsection{ Evaluation of ${\cal N}^{\rm E}_{zz}(\omega)$}

We now place the electric dipole moment along the $z$-axis $\bm p_0 = p_0 \hat{z}$, in which case the Maxwell equations near $z = z_{\rm q}$ read:
\begin{align}
    \nabla \cdot \bm D = 0,\quad \nabla \cdot \bm B = 0,\quad
    \nabla\times(\bm D - \bm P) + \varepsilon_0 \partial_t \bm B = 0,\quad
    \nabla \times \bm B - \mu_0 \partial_t \bm D = 0, \label{eqn:Maxwell_Disp}
\end{align}
where $\bm D = \varepsilon_0 \bm E + \bm P$ is the displacement field. From Eqs.~\eqref{eqn:Maxwell_Disp}, we obtain the boundary conditions at $z = z_{\rm q}$:
\begin{align}
    E_z^+ = E_z^-,\quad B_z^+ = B_z^-,\quad
    E_\perp^+ = E_\perp^-,\quad
    E_\parallel^+ - E_\parallel^- = - i q p_0/\varepsilon_0,\quad \bm B^+_t = \bm B^-_t.
\end{align}
After similar algebra as above, we arrive at:
\begin{align}
     {\cal N}^{\rm E}_{zz}(\omega) \approx \frac{T}{64\pi\omega^2\varepsilon_0^2 z_{\rm q}^4} \int_0^\infty dx \, x^3 \, e^{-x} \, \text{Re}\Big[ \frac{\tilde{\sigma}_{\rm L}(x/(2z_{\rm q}),\omega)}{\varepsilon(x/(2z_{\rm q}),\omega) } \Big]\approx2 {\cal N}^{\rm E}_{xx}(\omega).
\end{align}
Finally, the non-local optical conductivity should be expressed in terms of the phonon Green's function via the equations provided in Appendix ~\ref{appendix:sigma+D}.

\subsection{$1/T_2$-measurements and the electric noise}

An interesting aspect of spin-1 qubits (like NV centers) is that electric field fluctuations can also contribute to the spin-echo signal~\cite{myers2017double}. In the WC phase,  the electric noise is much stronger than the magnetic one. However, the coupling of the electric field to the qubit is expected to be significantly weaker than the magnetic coupling~\cite{myers2017double}. 
The net effect for NV centers is that the correction to $T_2$ from electric field fluctuations is expected to be comparable to the magnetic $T_2$-time. 
The resulting $T_2$-time, therefore, is expected to be appreciably shorter than the one considered above in Appendix~\ref{appendix:SER}, which should facilitate the feasibility of spin-echo-like measurements.  
Careful analysis of such electric-field correction can be done using the framework developed in the present paper. We also remark that it is conceivable that for other spin-1 detectors, the electric field coupling can be bigger so that the electric noise  for WCs might dominate the spin-echo-like signal.

\section{Phonon dispersion of a weakly-coupled clean bilayer Wigner crystal}
\label{appendix:dispersion}

In this Appendix, we derive analytic expressions for the dispersion relation of a clean bilayer WC at small-$q$ and in the limit of weak interlayer coupling, relevant for the TMD bilayer WCs recently realized in Ref.~\cite{Zhou_2021}.
The dispersion relations of the clean crystal derived in this Appendix may then be used as input for calculations that treat disorder effects -- Appendix~\ref{appendix:GVM}.

The potential energy of a bilayer electron system is
    \be
    U = U_1 + U_2 + U_{12},
    \ee
where $U_{1}$ and $U_2$ are the potential energies of individual layers and $U_{12}$ is the interlayer potential energy:
    \begin{align}
    U_i = \frac 12 \sum_{\bm r_i \neq \bm r'_i} \frac{e^2}{|\bm r_i - \bm r'_i|} \quad (i = 1,2), \qquad
    U_{12} = \sum_{\bm r_1, \bm r_2} \frac{e^2}{\sqrt{|\bm r_1 - \bm r_2|^2 + d^2}}.
    \end{align}
Here $\bm r_1$ and $\bm r_2$ refer to coordinates of electrons in layers 1 and 2, respectively. The phonon spectrum is obtained by writing $\bm r_i = \bfR_i + \bm u_i(\bfR_i)$, where $\bfR_i$ label the equilibrium lattice sites in layer $i$, and expanding the potential energy to the quadratic order in $\bm u$. Carrying this out for $U_{1,2}$ yields the phonon spectrum of individual layers:
    \be
    \delta U_i = \frac 12 \sum_{\bm q \in \rm{1BZ}} \left[ m\omega_{\rm{L}}^2(\bm q) u_{{\rm L}, i}(\bm q) u_{{\rm L}, i}(-\bm q) + m\omega_{\rm{T}}^2(\bm q) u_{{\rm T}, i}(\bm q) u_{{\rm T}, i}(-\bm q) \right], \quad (i = 1, 2).
    \label{eq:decoupled_spec}
    \ee
The frequency dispersion curves for the longitudinal ($\omega_{\rm{L}}$) and transverse ($\omega_{\rm{T}}$) branches have been calculated in Ref.~\cite{bonsall_elastic_wigner}. Here we record the long-wavelength behavior:
    \be
    \omega_{\rm L}^2(\bm q) = \omega_p^2 (qa) - \omega_p^2 \alpha_0 (qa)^2, \qquad 
    \omega_{\rm T}^2(\bm q) = \omega_p^2 \beta_0 (qa)^2,
    \ee
where $a$ is the WC lattice constant, $\omega_p^2 a = 2\pi e^2 / (m A_c)$ with $A_c$ the unit-cell area, and the numerical coefficients are $\alpha_0 \approx 0.181483$ and $\beta_0 \approx 0.0362967$.

\begin{figure}
    \centering
    \includegraphics{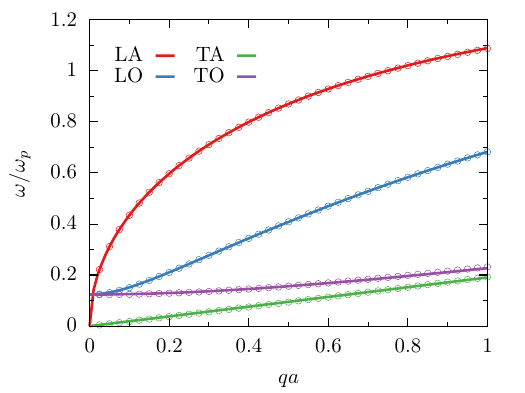}
    \caption{Dispersion curves for the bilayer WC for interlayer spacing $d/a=1$, representative of the weak interlayer coupling limit. Solid lines are the analytic small-$q$ approximation from Eqs. \eqref{eq:LA}-\eqref{eq:TO}. Open symbols are the exact numerical dispersion curves computed as in Ref.~\cite{Goldoni:1996}.}
    \label{fig:bilayer_disp}
\end{figure}

We now turn to the interlayer term. The quadratic change to the energy is
    \be
   \delta U_{12} = -\frac 12 \sum_{\bfR_1, \bfR_2} \phi_{\alpha \beta}(\bfR_1 - \bfR_2) [u_1^\alpha(\bfR_1) - u_2^\alpha(\bfR_2)][u_1^\beta(\bfR_1) - u_2^\beta(\bfR_2)], \qquad  \phi_{\alpha\beta}(\bm r) \equiv -\partial_\alpha \partial_\beta \frac{e^2}{\sqrt{r^2 + d^2}}.
   \label{eq:quad_U12}
    \ee
In the case of two weakly coupled layers, the lowest energy structure corresponds to two staggered triangular lattices, displaced relative to each other by a vector $\bfc$ \cite{Goldoni:1996}. With this in mind, we Fourier transform Eq.~\eqref{eq:quad_U12}:
    \be
    \delta U_{12} = \sum_{\bm q \in \rm{1BZ}} \tilde \phi_{\alpha\beta}(\bm q) u_1^\alpha(\bm q) u_2^\beta(-\bm q) - \frac 12 \sum_{\bm q \in \rm{BZ}} \tilde \phi_{\alpha\beta}(\bm q = 0) [ u_1^\alpha(\bm q) u_1^\beta(-\bm q) + u_2^\alpha(\bm q) u_2^\beta(-\bm q) ],
    \label{eq:quad_U12q}
    \ee
where 
    \be
    \tilde \phi_{\alpha \beta}(\bm q) = \sum_\bfR e^{-i\bm q \cdot \bfR} \phi_{\alpha\beta}(\bfR - \bfc).
    \ee
Here the sum is  taken over vectors $\bfR$ on the triangular lattice. Function $\tilde \phi_{\alpha\beta}(\bm q)$ may also be written as a sum over reciprocal lattice vectors $\bfG$:
    \begin{align}
    \tilde \phi_{\alpha\beta}(\bm q) &= \frac{2\pi e^2}{A_c}\sum_\bfG  e^{-i (\bm q - \bfG)\cdot \bfc} (\bm q - \bfG)_\alpha (\bm q -\bfG)_\beta \frac{ e^{-|\bm q - \bfG|d}}{|\bm q - \bfG|} \notag\\
    &= \frac{2\pi e^2}{A_c} e^{-i \bm q \cdot \bfc} \frac{q_\alpha q_\beta}{q} e^{-q d} + \frac{2\pi e^2}{A_c}\sum_{\bfG \neq 0}  e^{-i (\bm q - \bfG)\cdot \bfc} (\bm q - \bfG)_\alpha (\bm q -\bfG)_\beta \frac{ e^{-|\bm q - \bfG|d}}{|\bm q - \bfG|}.
    \end{align}
In the limit of weak interlayer coupling (large spacing $d$), the second term is well approximated by setting $\bm q = 0$ and summing over  the first shell of reciprocal lattice vectors only. This yields
    \be
    \tilde \phi_{\alpha\beta}(\bm q) \approx \frac{2\pi e^2}{A_c} e^{-i \bm q \cdot \bfc} \frac{q_\alpha q_\beta}{q} e^{-q d} - \frac{2\pi e^2}{A_c} \frac{1}{a} 2\pi \sqrt 3 e^{-\frac{4\pi}{\sqrt 3} \frac da}  \delta_{\alpha\beta} \equiv  m\omega_p^2 (qa) e^{-qd} \frac{q_\alpha q_\beta}{q^2} - \frac 12 m\omega_{\rm{opt}}^2 \delta_{\alpha\beta}.
    \ee
Equation~\eqref{eq:quad_U12q} then gives
    \be
    \begin{aligned}
    \delta U_{12} &\approx \sum_{\bm q \in \rm{BZ}} \left\{ m\omega_p^2 qa e^{-qd} u_{\rm{L}, 1}(\bm q) u_{\rm{L}, 2}(-\bm q) - \frac 12 m\omega_{\rm{opt}}^2 [ u_{\rm{L}, 1}(\bm q) u_{\rm{L}, 2}(-\bm q) + u_{\rm{T}, 1}(\bm q) u_{\rm{T}, 2}(-\bm q)] \right\} \\
    &\quad + \frac 12 \sum_{\bm q \in \rm{BZ}} \frac 12 m\omega_{\rm{opt}}^2 [u_{\rm{L}, 1}(\bm q) u_{\rm{L}, 1}(-\bm q) + u_{\rm{T}, 1}(\bm q) u_{\rm{T}, 1}(-\bm q) + u_{\rm{L}, 2}(\bm q) u_{\rm {L}, 2}(-\bm q) + u_{\rm{T}, 2}(\bm q) u_{\rm{T}, 2}(-\bm q)].
    \end{aligned}
    \ee
Combining this with Eq.~\eqref{eq:decoupled_spec}, we obtain the dynamical matrix for the weakly coupled bilayer system:
    \be
    \begin{aligned}
    \delta U = \frac 12 m\sum_{\bm q \in \rm{BZ}} &\left[ \begin{pmatrix} u_{\rm{L}, 1}(\bm q) &  u_{\rm{L}, 2}(\bm q)\end{pmatrix} 
    \begin{pmatrix}
    \omega_{\rm L}^2(\bm q) + \frac 12 \omega_{\rm{opt}}^2 & \omega_p^2 (qa) e^{-qd} -\frac 12 \omega_{\rm{opt}}^2 \\[1em]
    \omega_p^2 (qa) e^{-qd} -\frac 12 \omega_{\rm{opt}}^2 & \omega_{\rm L}^2(\bm q) + \frac 12 \omega_{\rm{opt}}^2
    \end{pmatrix}
    \begin{pmatrix} u_{\rm{L}, 1}(-\bm q) \\  u_{\rm{L}, 2}(-\bm q)\end{pmatrix} \right. \\[2 em]
    &\quad \left. +  \begin{pmatrix} u_{\rm{T}, 1}(\bm q) &  u_{\rm{T}, 2}(\bm q)\end{pmatrix} 
    \begin{pmatrix}
    \omega_{\rm T}^2(\bm q) + \frac 12 \omega_{\rm{opt}}^2 & \omega_p^2 (qa) e^{-qd} -\frac 12 \omega_{\rm{opt}}^2 \\[1 em]
    \omega_p^2 (qa) e^{-qd} -\frac 12 \omega_{\rm{opt}}^2 & \omega_{\rm T}^2(\bm q) + \frac 12 \omega_{\rm{opt}}^2
    \end{pmatrix}
    \begin{pmatrix} u_{\rm{T}, 1}(-\bm q) \\  u_{\rm{T}, 2}(-\bm q)\end{pmatrix}  \right].
    \end{aligned}
    \ee
Diagonalizing the system yields four phonon branches: longitudinal acoustic (LA), longitudinal optical (LO), transverse acoustic (TA), and transverse optical (TO), with dispersions
    \begin{align}
	\omega_{\rm{LA}}^2(\bm q) &= \omega_p^2(1 + e^{-qd}) (q a) - \omega_p^2\alpha_0 (qa)^2, \label{eq:LA} \\
	\omega_{\rm{LO}}^2(\bm q) &= \omega_{\rm{opt}}^2 + \omega_p^2(1 - e^{-qd}) (q a)  - \omega_p^2\alpha_0 (q a)^2, \label{eq:LO} \\
	\omega_{\rm{TA}}^2(\bm q) &= \omega_p^2 \beta_0 (qa)^2, \label{eq:TA} \\
	\omega_{\rm{TO}}^2(\bm q) &= \omega_{\rm{opt}}^2 + \omega_p^2 \beta_0 (qa)^2. \label{eq:TO}
	\end{align} 
The optical phonon frequency is given by:
    \be
	\omega_{\rm{opt}}^2 = \omega_p^2 (4\pi\sqrt 3) 
 \exp\Big\{-\frac{4\pi}{\sqrt 3} \frac da \Big\}.
	\ee
Comparisons between the approximate small-$q$ dispersion in Eqs.~\eqref{eq:LA}-\eqref{eq:TO} and the exact numerical dispersion curves, computed as in Ref.~\cite{Goldoni:1996}, are given in Fig.~\ref{fig:bilayer_disp}.

\section{Gaussian variational treatment of disordered bilayer Wigner crystals}
\label{appendix:GVM}

In this Appendix, we present the analysis of disordered bilayer Wigner crystals, which are treated using the replica trick and the framework of the Gaussian variational approach developed in Refs.~\cite{giamarchi_ledoussal_1995_PhysRevB52_1242, Giamarchi:1995, chitra_wigner_hall, chitra_wigner_zerob}. We refer the reader to those references for technical details and here only summarize the essential ingredients. Treating disorder via the replica trick yields the following imaginary-time action:
\begin{align}
    {\cal S} = {\cal S}_0 + {\cal S}_{\rm dis}, \label{eqn:S_full_bl}
\end{align}
where
\begin{align}
    {\cal S}_0 = \frac{1}{2} \sum_{n,\bm q,\lambda,a} u_\lambda^a(-\bm q,-i\omega_n) {\cal D}_{0,\lambda}^{-1}(\bm q,i\omega_n) u_\lambda^a (\bm q,i\omega_n),
    \label{eq:S0}
\end{align}
and
\begin{align}
    {\cal S}_{\rm dis} = - \frac{\rho_0^2}{2} \sum_{l= 1,2}  \int_{0}^\beta d\tau \int_{0}^\beta d\tau'\int d^2\bm r \sum_{a,b} \sum_{\bm G}  
    \Delta_{\bm G} \cos \{ \bm G \cdot (\bm u^a_l(\bm r,\tau) - \bm u^b_l(\bm r,\tau') )\}.
    \label{eq:Sdis}
\end{align}
The imaginary-time action~\eqref{eqn:S_full_bl} represents the bilayer generalization of Eq.~(12) of Ref.~\cite{chitra_wigner_zerob}. Here $a$ is the replica index, $\omega_n = 2\pi n T$ are the bosonic Matsubara frequencies, $\Delta_{\bf G}$ are the Fourier components of the disorder correlation function, with $\bfG$ being the WC reciprocal lattice vectors. The bare phonon propagator is given by:
\begin{align}
    {\cal D}_{0,\lambda}(\bm q, i\omega_n) = \frac{1}{m} \frac{1}{\omega_n^2 + \omega_\lambda^2(\bm q)}.
\end{align}
The index $\lambda = \{\text{LA, LO, TA, TO}\}$ refers to the four eigenmodes of the elastic Hamiltonian of a bilayer WC: Label L (T) stands for a longitudinal (transverse) mode, and A (O) denotes an acoustic (optical) mode. Provided there is no time-reversal symmetry breaking (here we consider the case with no applied magnetic field), one can write:
\begin{gather}
    u_{\alpha,l}(\bm q,i\omega_n) = \frac{1}{\sqrt{2}} \Big[ u_{\rm LA}(\bm q,i\omega_n) + (-1)^l u_{\rm LO}(\bm q,i\omega_n) \Big] \hat{q}_\alpha +  \frac{1}{\sqrt{2}} \Big[ u_{\rm TA}(\bm q,i\omega_n) + (-1)^l u_{\rm TO}(\bm q,i\omega_n) \Big] \varepsilon_{\alpha\beta} \hat{q}_\beta,
\end{gather}
where $l = 1,2$ is the layer index.

\subsection{Gaussian variational method}

The Gaussian variational method approximates the full nonlinear action with the best trial quadratic action:
\begin{align}
    {\cal S}_{\rm tr} [D] =  \frac{1}{2} \sum_{n,\bm q} \sum_{ab}\sum_{l l^{'}}\sum_{\alpha\alpha{'}} u_{\alpha l}^a(-\bm q,-i\omega_n) (D^{-1})_{\alpha l,\alpha' l^{'}}^{ab} (\bm q,i\omega_n) u_{\alpha' l^{'}}^b (\bm q,i\omega_n),
\end{align}
where the phonon propagator $D$ is a variational quadratic form.  The variational free energy is
\begin{align}
    {\cal F}_{\rm var} = {\cal F}_{\rm tr} + 
  T\langle {\cal S} - {\cal S}_{\rm tr} \rangle_{\rm tr}.
\end{align}
The saddle-point equation, which determines $D$, then  reads:
\begin{align}
    \frac{\delta {\cal F}_{\rm var} }{\delta D_{\alpha l,\alpha' l^{'}}^{ab} (\bm q,i\omega_n) } = 0.
\end{align}
Explicit evaluation of this variational derivative gives the Dyson equation:
\begin{align}
    (D^{-1})_\lambda^{ab} (\bm q,i\omega_n) = \delta^{ab} {\cal D}^{-1}_{0,\lambda}(\bm q,i\omega_n) - {\Pi}^{ab}(i\omega_n),\label{eqn:Dyson_1}
\end{align}
where
\begin{align}
    \Pi^{ab}(i\omega_n)  = & -\delta^{ab} \frac{\rho_0 }{2}  \int_{0}^\beta d\tau \sum_{\bm G}  
    \Delta_{\bm G} G^2\Big( \sum_c e^{-\frac{1}{2}G^2B^{ac}(\tau)} - 
    \cos(\omega_n\tau) e^{- \frac{1}{2} G^2 B^{aa}(\tau)}
    \Big)  \notag\\
    & 
    + (1-\delta^{ab})\frac{\rho_0 }{2}  \int_{0}^\beta d\tau \sum_{\bm G}  
    \Delta_{\bm G} G^2 \cos(\omega_n\tau) e^{-\frac{1}{2}G^2B^{ab}(\tau)},\label{eqn:Dyson_2}\\
    B^{ab}(\tau) = & \frac{1}{4\beta N} \sum_{\bm q,n,\lambda} \Big( D^{aa}_\lambda(\bm q,i\omega_n) + D^{bb}_\lambda(\bm q,i\omega_n) - 2 \cos(\omega_n\tau)D^{ab}_\lambda(\bm q,i\omega_n) 
    \Big).\label{eqn:Dyson_3}
\end{align}
Here $N$ is the total number of lattice sites. Equations~\eqref{eqn:Dyson_1}-\eqref{eqn:Dyson_3} are to be solved self-consistently in the limit where the total number of replicas goes to zero, $n \to 0$.

\subsection{Generic structure of  replica symmetry broken solutions }

The $n \to 0$ limit is understood within the standard algebra of replica matrices \cite{Giamarchi:1995, chitra_wigner_long, mezard_parisi_1991_replica_JournPhysI1_809}: Diagonal matrix elements are replaced according $D^{aa}\to \Tilde{D}$ and off-diagonal components are parameterized by a continuous variable $0< u < 1$: $D^{a\neq b}\to D(u)$. The self-energy in Eq.~\eqref{eqn:Dyson_2} is then written as:
\begin{gather}
    \Tilde{\Pi} (i\omega_n) = -\frac{\rho_0 }{2}  \int_{0}^\beta d\tau \sum_{\bm G}  
    \Delta_{\bm G} G^2\left( [1 - \cos(\omega_n\tau)]e^{- \frac{1}{2} G^2 \Tilde{B}(\tau)} -  \int_0^1 du\,  e^{-\frac{1}{2}G^2B(\tau,u)} 
    \right),
    \\
    \Pi(i\omega_n,u)  =  \frac{\rho_0 }{2}  \int_{0}^\beta d\tau \sum_{\bm G}  
    \Delta_{\bm G} G^2 \cos(\omega_n\tau) e^{-\frac{1}{2}G^2B(\tau,u)}.
\end{gather}
The fact that we consider static disorder with no temporal correlations implies that off-diagonal matrix elements are $\tau$-independent, so that $\Pi(i\omega_n,u) = \delta_{n,0} \Pi(u)$, where
\begin{align}
    \Pi(u) = \frac{\beta \rho_0}{2} \sum_{\bm G}  
    \Delta_{\bm G} G^2 e^{-\frac{1}{2}G^2B(u)}. 
    \label{eqn:Pi_u_v0}
\end{align}

Following Refs.~\cite{Giamarchi:1995,chitra_wigner_long}, a solution to the Dyson equation~\eqref{eqn:Dyson_1} with replica symmetry breaking (RSB) has the property:
\begin{equation}
   [\Pi](u) =  m\omega_0^2 \times 
    \begin{cases}
   f(u), & 0< u < u_c\\
   1, & u_c\leq u < 1
 \end{cases},\label{eqn:RSB_gen}
\end{equation}
where we have defined $[A](u)\equiv uA(u) - \displaystyle\int_0^u dv\,A(v)$. Here $f(u)$ is a dimensionless function with $f(u_c) = 1$. At this stage, the parameters $f(u)$, $u_c$, and $\omega_0$ are yet unknown and will be self-consistently determined below. By using the form~\eqref{eqn:RSB_gen} and the inversion formulas for replica matrices (see Appendix II of Ref.~\cite{mezard_parisi_1991_replica_JournPhysI1_809}), one can rewrite the Dyson equation~\eqref{eqn:Dyson_1} as:
\begin{gather}
   D_{c,\lambda}^{-1}(\bm q,i\omega_n)   = {\cal D}^{-1}_{0,\lambda}(\bm q,i\omega_n) + P(i\omega_n) + (1 - \delta_{n,0}) m\omega_0^2,
   \label{eqn:RSB_in}
   \\
    P(i\omega_n)  =  \frac{\rho_0}{2}  \int_{0}^\beta d\tau \sum_{\bm G}  
    \Delta_{\bm G} G^2 [1 - \cos(\omega_n\tau)] \left( e^{- \frac{1}{2} G^2 \Tilde{B}(\tau)} -   e^{-\frac{1}{2}G^2B(u_c)} 
    \right),\\
    B(u_c)  =  \frac{1}{2\beta N} \sum_{\bm q,n \neq 0,\lambda} {D}_{c,\lambda}(\bm q,i\omega_n ) + \frac{1}{2\beta N} \sum_{\bm q,\lambda} \frac{1}{ {\cal D}_{0,\lambda}^{-1}(\bm q,i\omega_n = 0) + m\omega_0^2 },\\
     \Tilde{B}(\tau)   =  \frac{1}{2\beta N} \sum_{\bm q,n,\lambda} (1 - \cos(\omega_n\tau)) D_{c,\lambda}(\bm q,i\omega_n),\label{eqn:RSB_fin}
\end{gather}
where $D_{c,\lambda}(\bm q,i\omega_n) \equiv \Tilde{D}_\lambda(\bm q,i\omega_n) - \langle D_\lambda(\bm q,i\omega_n)\rangle$ is the connected correlation function and $\langle A \rangle \equiv \displaystyle\int_0^1 du\,A(u)$.

Equations \eqref{eqn:RSB_in}-\eqref{eqn:RSB_fin} form a closed set of equations for the phonon Green's function, in which $\omega_0$ enters as a parameter. Furthermore, these equations are independent of the precise nature of the RSB (full, one-step, etc.) and the functional form of $\Pi[u]$. Function $\Pi[u]$ (and from it $\omega_0$) are determined separately in the next subsection.

\subsection{One-step RSB}

We limit ourselves to the so-called single-cosine approximation, where the summations over reciprocal lattice vectors $\bfG$ are truncated to the first `momentum shell' of six wave vectors with $\bfG = \bfG_1$:
\begin{align}
    \Pi(u) \approx \frac{\beta \rho_0 }{2}  \Delta G^2 e^{-\frac{1}{2}G^2B(u)}, 
\end{align}
where $\Delta  = 6 \Delta_{\bfG_1}$. In two spatial dimensions, the one-step RSB solution has the following structure \cite{Giamarchi:1995, chitra_wigner_long}:
\begin{equation}
   [\Pi](u) = 
    \begin{cases}
   0, & u < u_c\\
   m\omega_0^2, & u_c\leq u
 \end{cases},\quad \Pi(u) = 
    \begin{cases}
   0, & u < u_c\\
   m\omega_0^2/u_c, & u_c\leq u
 \end{cases}, \quad B(u) = 
    \begin{cases}
   \infty, & u < u_c\\
   B_c, & u_c\leq u
 \end{cases}.
\end{equation}
At the moment, we have three unknowns: $\omega_0$, $u_c$, and $B_c$. The value of $B_c$ can be determined by (numerically) solving Eqs.~\eqref{eqn:RSB_in}-\eqref{eqn:RSB_fin}. The remaining equations are:
\begin{align}
    m\omega_0^2 = \frac{u_c \beta \rho_0 }{2}  \Delta G^2 e^{-\frac{1}{2}G^2B_c} \label{eqn:Omega_sc}
\end{align}
and
\begin{align}
    \frac{\rho_0}{8}  
    \Delta G^4  e^{-\frac{1}{2}G^2B_c} \times \frac{1}{N}\sum_{\bm q,\lambda} \frac{1}{ ({\cal D}_{0,\lambda}^{-1}(\bm q,i\omega_n = 0) + m \omega_0^2 )^2} = 1. \label{eqn: B_c_sc}
\end{align}
The first one directly follows from the above definitions, while the second one is a bit subtle. It could be derived, for instance, by considering $d = 2+ \varepsilon$ (in which case the function $[\Pi](u)$ is continuous) and then taking the limit $\varepsilon \to 0$.

In practice, we solve Eqs.~\eqref{eqn:RSB_in}-\eqref{eqn:RSB_fin}, Eq.~\eqref{eqn:Omega_sc}, and Eq.~\eqref{eqn: B_c_sc} numerically, both in real and imaginary times. The former allows us to evaluate various experimentally relevant response functions but requires an additional step of analytical continuation $i\omega_n \to \omega + i0$, which is done by following step-by-step Appendix D of Ref.~\cite{Giamarchi:1995}. 

Finally, to obtain the optical conductivity shown in Fig.~\ref{fig:sig_bilayer}, we used a simplified phonon spectrum, consisting only of two branches:
    \be
    \omega_{\rm A}^2 = v_s^2 q^2, \quad \omega_{\rm O}^2 = v_s^2 q^2 + \omega_{\rm{opt}}^2.
    \ee

\end{document}